\documentclass[pre,aps,superscriptaddress]{revtex4}
\usepackage{epsfig}
\usepackage{epsf}
\usepackage{graphicx}
\usepackage{dcolumn}
\usepackage{subfigure}
\usepackage{bm}
\usepackage{amssymb,latexsym,amsmath,dsfont}
\usepackage[dvips]{color}
\usepackage[latin1]{inputenc}

\def\be{\begin{equation}}
\def\ee{\end{equation}}
\def\ba{\begin{array}}
\def\ea{\end{array}}
\def\bea{\begin{eqnarray}}
\def\eea{\end{eqnarray}}

\begin{document}

\title{Avalanches, precursors and finite size fluctuations
\\ in a mesoscopic model of amorphous plasticity}

\author{Mehdi Talamali}
\affiliation{
Laboratoire PMMH, UMR 7636 CNRS/ESPCI/Univ. Paris 6 UPMC/Univ. Paris 7 Diderot\\
10 rue Vauquelin, 75231 Paris cedex 05, France}
\author{Viljo Pet\"aj\"a}
\affiliation{
Laboratoire SVI, UMR 125 CNRS/Saint-Gobain\\
39 quai Lucien Lefranc, 93303 Aubervilliers cedex 05, France}
\author{Damien Vandembroucq}
\affiliation{
Laboratoire PMMH, UMR 7636 CNRS/ESPCI/Univ. Paris 6 UPMC/Univ. Paris 7 Diderot\\
10 rue Vauquelin, 75231 Paris cedex 05, France}
\author{St\'ephane Roux}
\affiliation{ LMT-Cachan, ENS de Cachan/CNRS-UMR 8535/Univ. Paris 6 UPMC/PRES
    UniverSud Paris\\
61 Avenue du Pr\'esident Wilson - 94235 Cachan cedex, France}

\begin{abstract}

We discuss avalanche and finite size fluctuations in a mesoscopic
model 
to describe the shear plasticity of amorphous materials.  Plastic deformation is
assumed to occur through series of local reorganizations. Yield stress criteria
are random while each plastic slip event induces a quadrupolar long range
elastic stress redistribution.  The model is discretized on a regular square
lattice.
Shear plasticity can be studied in this context as a depinning dynamic phase
transition. We show evidence for a scale free distribution of avalanches
$P(s)\propto S^{-\kappa}$ with a non trivial exponent $\kappa \approx 1.25$
significantly different from the mean field result $\kappa = 1.5$. Finite
size effects allow for a characterization of the scaling invariance of the yield
stress fluctuations observed in small samples. We finally identify a population
of precursors of plastic activity and characterize its spatial distribution. \\
\end{abstract}

\date{\today}

\maketitle

\section{Introduction}

While traditionally described in continuum mechanics by constitutive laws at
macroscopic scale, it has progressively appeared in the last two decades that
the mechanical behavior of materials was not as smooth and regular as
anticipated. In particular crack propagation in brittle materials and plastic
flow in crystalline solids have been shown to exhibit jerky motion and scale
free spatio-temporal
correlations~\cite{Miguel-Nat01,Zaiser-AdvPhys06,Bonamy-JPhysD09}.

Beyond its obvious fundamental interest, the understanding of
intermittence and intrinsic fluctuations in mechanics of materials and
their consequences at macroscopic scale is of direct importance for
engineering applications: among other examples a quantitative
assessment of risk of failure would allow to better determine security
margins which stay rather uncontrolled (and often overestimated); a
theoretical understanding of finite size effects would allow to model
the mechanical behavior of small pieces, a question of crucial
interest with the rapid technological development of MEMS and NEMS
(Micro or Nano- Electro-Mechanical Systems).

Concepts such as avalanches and criticality have thus been growingly used in
that context to describe and model fracture and plasticity. In particular, the
paradigm of the depinning transition has shown extremely appealing to model such
non-linear phenomena at mesoscopic
scale~\cite{Schmittbuhl-PRL95,Daguier-EPL95,Ramanathan-PRL97b,Fisher-PR98,Zaiser-MSEA01,BVR-PRL02,Moretti-PRB04}.
Such a formalism indeed naturally captures the competition between the disorder
of local thresholds (toughness for crack propagation, yield stress for plastic
deformation) and elastic interactions which couple local mechanical events
(crystallographic slip, crack advance). A critical threshold naturally emerges
at macroscopic scale which separates a static phase (crack propagation or
plastic deformation stops after a finite excursion) from a mobile phase (free
propagation or deformation). As usual, this dynamic phase transition can be
characterized by a set of critical exponents.

While crystalline plasticity or crack propagation rely on rather solid
grounds (theory of dislocation and linear elastic fracture mechanics
respectively), the understanding of plastic deformation in amorphous
materials such as oxide or metallic glasses is still in its
infancy. In absence of crystalline lattice, plasticity seems to
originates from a series of very local structural
rearrangements~\cite{Argon-ActaMet79,FalkLanger-PRE98}.  Beyond this
first level of description, any local reorganization has to be
accommodated by the surrounding elastic matrix, and induces internal
stress~\cite{Maloney-PRL04a,Tanguy-EPJE06,Rodney-PRL09}. These local
plastic events thus do not occur independently but in a strong
correlated way.

We recently introduced a mesoscopic model of plasticity in amorphous
materials~\cite{TPRV-Meso09a}. Following an earlier
work~\cite{BVR-PRL02} we developed a scalar discrete model on a
regular lattice with a random yield stress. The local slip occurring
when the shear stress satisfies the plastic criterion is accompanied
by an elastic stress redistribution of quadrupolar
symmetry~\cite{Picard-EPJE04,TPRV-PRE08} which corresponds to the
elastic response of the surrounding matrix to this Eshelby-like
plastic inclusion\cite{Eshelby57}.  Although original due to the
quadrupolar symmetry of the elastic interaction, one recognizes in
this short description the two ingredients of a depinning model: a
random threshold field and an elastic interaction.

In a previous paper~\cite{TPRV-Meso09a} we focussed on the competition between
localization and diffusion which natural emerges from the peculiar symmetry of
the elastic interaction. Some directions being favored, plastic deformation
forms shear bands which span the entire lattice. This localization is however
not persistent and after they grow up to the size of the system, shear bands
tend to diffuse throughout the lattice. In particular we could make evidence for
anisotropic strain correlations which are strikingly similar to those recently
observed in an atomistic study of a binary Lennard-Jones glass under
compression~\cite{Maloney-JPCM08,Maloney-PRL09}.

In the present paper, a particular focus is given to the critical
properties of the model. We recall in Section~\ref{sec:model} the
definition of the model and its salient properties.  Avalanches are
quantitatively characterized in Section~\ref{sec:avalanche}.  Finite
size fluctuations close to critical point are analyzed in
Section~\ref{sec:finiteSize}. Characterization of avalanche precursor
sites is discussed in
Section~\ref{sec:precursor}. Section~\ref{sec:conclusion} concludes
this paper.

\begin{figure}[htbp]
 \subfigure[]{
  \label{fig:mini:subfig:a}
  \begin{minipage}[h]{0.45\textwidth}
   \centering
   \includegraphics[width=80mm]{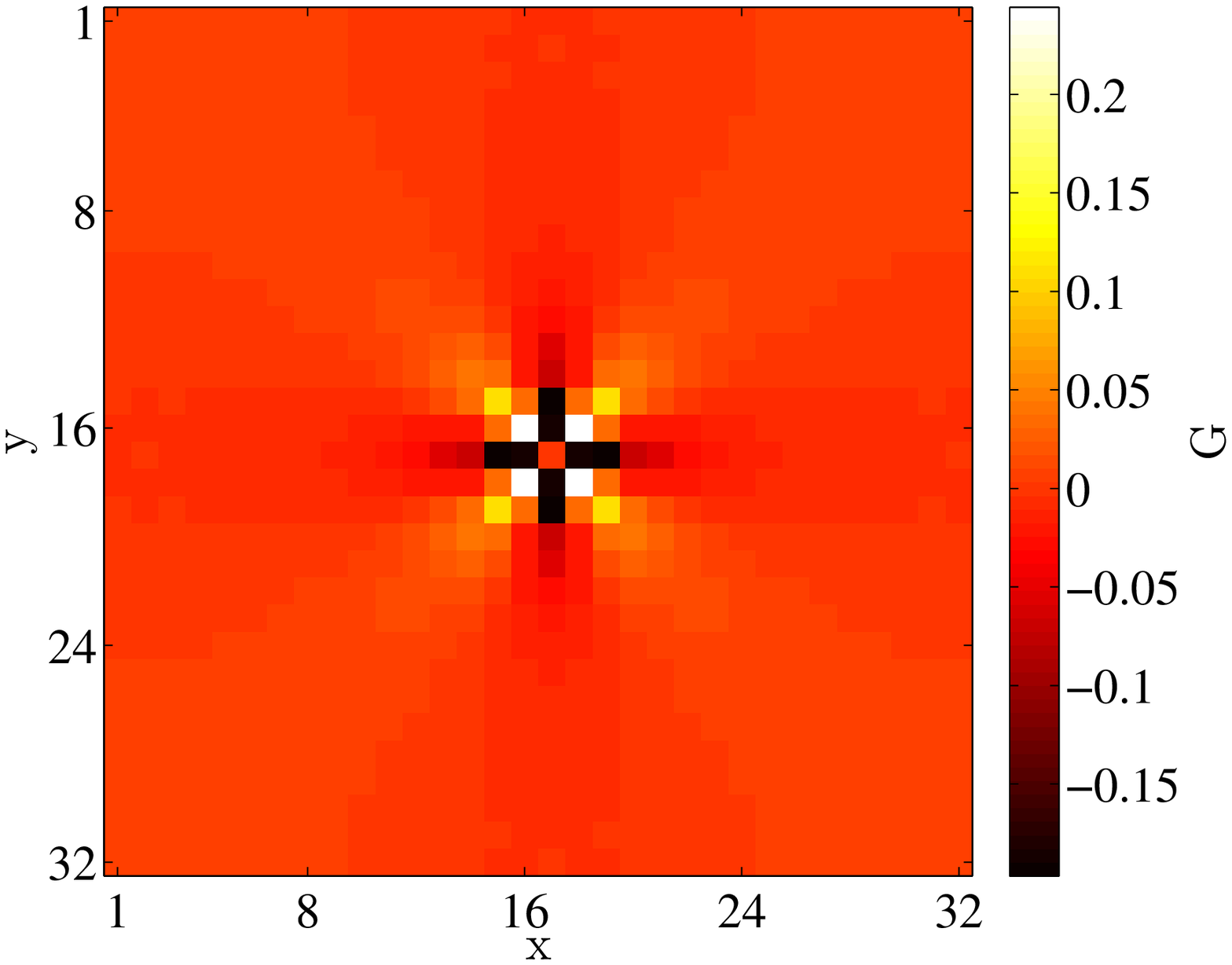}
  \end{minipage}
  \label{Green1}}
 \subfigure[]{
  \label{fig:mini:subfig:b}
  \begin{minipage}[h]{0.45\textwidth}
   \centering
   \includegraphics[width=77mm]{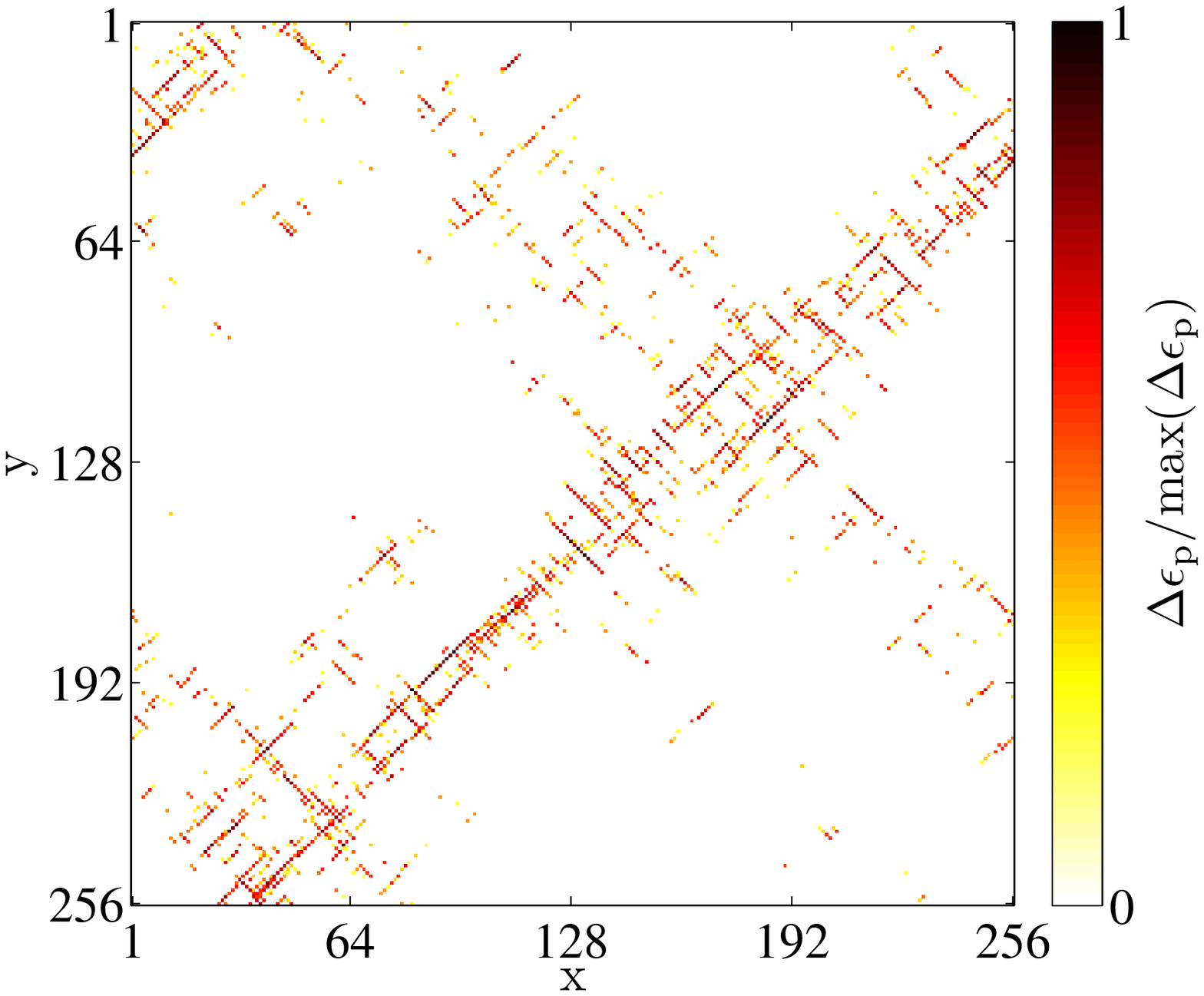}
  \end{minipage}
  \label{Green2}}
  \caption{(Color online) (a) Map of the quadrupolar elastic
    interaction $G$ used in the model. The discretization is performed in
    Fourier space and a inverse Fourier transform gives in the direct
    space the Green function satisfying the bi-periodic boundary
    conditions of the problem. (b) Map of cumulated plastic activity
    obtained for an averaged cumulated plastic strain $\Delta
    \varepsilon_p = 0.01$ taken at $\varepsilon_p=1.0$ (strain is
    expressed in arbitrary (infinitesimal) units). A clear
    localization of the plastic deformation is observed. Note that
    this localization behavior is non persistent (see
    Ref. ~\cite{TPRV-Meso09a} for details on the competition between
    localization and diffusion of the plastic deformation.)}
  \label{Fig1}
\end{figure}

\section{Brief description of the model}\label{sec:model}

A detailed description of the model can be found in ~\cite{TPRV-Meso09a}. Let us
simply summarize here the main points.

We consider an elastically homogeneous material in plane deformation geometry
under shear.  Discretization is performed on a square lattice with bi-periodic
boundary conditions at a scale which is larger than the size of a typical
rearrangement. This scale is to be large enough to allow the use of continuum
elasticity, and to neglect elastic inhomogeneities. At a large scale, we impose
a pure shear load $\sigma_{xx}=-\sigma_{yy}$, and $\sigma_{xy}=0$.

A simplification consists of assuming that local rearrangements induce an
elementary plastic shear with the same symmetry as the macroscopic imposed
macroscopic shear. We thus consider neither volumetric change, nor orientation
disorder for the shear principal axis at the microscopic scale. Consequently,
although the model is based on a genuine 2D elastic description, the tensorial
nature of the stress $\bm \sigma$ and strains $\bm \varepsilon$ plays no role.
Scalar (equivalent) stress $\sigma\equiv \sigma_{xx}-\sigma_{yy}$ and strain
$\varepsilon\equiv \varepsilon_{xx}-\varepsilon_{yy}$ can be defined. The latter
scalar stress (resp. strain) component will be called ``stress'' (resp.
``strain'') for simplicity in the following. The criterion for yielding
characterizes the local configuration of atoms, and hence will display some
variability.  A local yield threshold for each discrete site $\bm x$ as
$\sigma_{\gamma}(\bm x)$ is introduced, and will be treated as a random variable
in the sequel. For all sites, the same statistical distribution will be used,
chosen for simplicity as a uniform distribution over the interval $[0;1]$.  The
specific form of this distribution plays no role in the scaling features
addressed below.

The stress $\sigma$ is a sum of the externally applied stress
$\Sigma_{\mathrm{ext}}$ and the residual stress $\sigma_{\mathrm{res}}$ induced
by the previous rearrangements of other regions of the system. Thus, the local
scalar yield criterion for site $\bm x$ can be rewritten as
    \be
    \Sigma_{\mathrm{ext}}+\sigma_{\mathrm{res}}(\bm x) = \sigma_{\gamma}(\bm x).
    %\label{eq:criterion}
    \ee
where $\sigma_\gamma$ is the local yield stress. {Here and in the following we use upper (lower) case symbols for macroscopic (microscopic) quantities.}

Once this criterion is satisfied at site $\bm x$, the material experiences there
an incremental slip $\eta$ increasing the local plastic strain:
$\varepsilon_p(\bm x)\to\varepsilon_p(\bm x)+\eta$. Similarly to the yield
thresholds, the slip value $\eta$ is drawn randomly from the uniform
distribution, $[0;d]$ if not otherwise stated. As shown in Fig. \ref{Fig1}(a)
this local slip induces in turn a quadrupolar elastic stress
redistribution~\cite{Picard-EPJE04,TPRV-PRE08,TPRV-Meso09a}
$\sigma_{\mathrm{res}}(r,\theta)\to \sigma_{\mathrm{res}}(r,\theta) + \eta
\cos(4\theta)/r^2$.

In order to account for the local structural change that occurred the local
yield stress is renewed by drawing a new (uncorrelated) random value for
$\sigma_{\gamma}(\bm x)$. It is assumed that there is no persistence in the
local yield stress.

Quasi-static driving conditions are considered, using {\em extremal dynamics},
{\it i.e.} the imposed external loading $\Sigma_{\mathrm{ext}}$ is tuned at each
time step, $t$ at the current yield stress value, $\Sigma_c$, such that only one site can slip at a time:
    \be\ba{ll}
    \Sigma_{\mathrm{ext}}(t)&=\Sigma_c\\
    &=\min_{\bm x} [\sigma_{\gamma}(\bm x)-\sigma_{\mathrm{res}}(\bm x)]\\
    &=\sigma_{\gamma}(\bm x^*(t))-\sigma_{\mathrm{res}}(\bm x^*(t))
    \ea\ee
where $\bm x^*(t)$ is the extremal site at time $t$. Note however, that
``time'', $t$, is used here as a simple way of counting and ordering events.  On
average, time is simply proportional to the total plastic strain imposed on the
system, $\langle\epsilon_p\rangle=tdL^{-2}/2$. The plastic strain field is thus
simply \be \epsilon_p(\bm x,t)=\sum_1^t \eta(t)\delta(\bm x-\bm x^*(t)) \ee

%{\bf A few results stress vs strain, fluctuation of plastic strain}

In Ref.~\cite{TPRV-Meso09a}, we discussed the mechanical behavior of this model
and in particular we could make evidence for anisotropic plastic strain
correlations, signalling the formation of shear bands as illustrated in
Fig.\ref{Fig1}(b) which however were not persistent, and diffused throughout the
system over long times.  Under application of shear, a transient hardening stage
was observed before the shear stress eventually saturates. This original
phenomenon in the context of amorphous materials (in crystalline material
hardening is usually associated with dislocation pinning by impurities or
dislocation interactions) was interpreted as a consequence of the progressive
exhaustion of the weakest sites of the system (reminiscent of self-organized
critical systems). Plastic strain fluctuations were shown to exhibit a non
trivial scaling : its standard deviation $\rho(\varepsilon_p)$ grows as
$\rho(\varepsilon_p)\propto \varepsilon_p^\alpha $ with $\alpha\approx 0.75$ in
the transient regime and, in the stationary regime, the power spectrum of
plastic strain was shown to exhibit an anisotropic scaling $S(q,\theta) \propto
a(\theta) q^{-\alpha(\theta)} $ with $\alpha(\theta) $ obeying a quadrupolar
like symmetry. In particular, in the direction of the shear bands, we obtained
$\alpha_{\pi/4} \approx 1.7$.

\section{Avalanche behavior}\label{sec:avalanche}

As discussed above, while intermittence and avalanches were first identified in
earthquake dynamics, biological evolution~\cite{Bak-book96} or
magnetism~\cite{Sethna-PRL93,Zapperi-PRB98}, the recent years have also shown
their interest in the framework of mechanics of materials.

%{\bf Experimental results}
In the context of plasticity of crystalline materials, a significant amount of
results have been obtained over the last decade (see e.g. the comprehensive
review by M. Zaiser about scale invariance in plastic
flow~\cite{Zaiser-AdvPhys06}).  Acoustic emission measurements performed on ice
or metal monocrystals have shown a power law distribution of the energy
$P(E)\propto E^{-\kappa}$ with $\kappa\approx 1.6$ for
ice~\cite{Richeton-ActaMat05} and $\kappa\approx 1.5$ for hcp metals and
alloys~\cite{Richeton-MSEA06}.  The case of polycrystal is somewhat more complex
since not only a grain size related cut-off appears in the avalanche
distribution but the power law exponent is also significantly
lowered~\cite{Richeton-ActaMat05}. Performing nano-indentation measurements on
Nickel monocrystals, Dimiduk {\it et al} made evidence for a scale-free
intermittent plastic flow and estimated $\kappa \approx
1.5-1.6$~\cite{Dimiduk-Science06}.

Very recently analogous analysis could be performed on metallic glass samples.
Sun {\it et al}~\cite{Wang-PRL10} measured the distributions of stress drops
occurring in the strain stress curve were for various metallic glass samples
under compression. Scale free distributions were observed with a power law
exponent $\kappa \in [1.37-1.49]$.

%
%\subsection{Numerical results}

Various models have been designed which capture this avalanche
behavior in plasticity at least from a qualitative point of
view. Dislocation dynamics~\cite{Miguel-Nat01} and phase
field~\cite{Zaiser-AdvPhys06} models have for instance been used in
that purpose in the case of crystal plasticity. In the same context,
Moretti and Zaiser~\cite{Zaiser-JSM05,Zaiser-AdvPhys06} developed at
mesoscopic scale a model very similar to the one presented here since
it integrates some local yield randomness. A significant difference
stems from their account of short range interaction between
dislocations moving on close slip planes. This local elastic
interaction thus adds up and competes with the long range interaction
ensuring compatibility.  This model was then used to analyze slip
avalanches in crystal plasticity~\cite{Zaiser-JSM07}, a scale free
behavior was obtained with a power law exponent $\kappa =1.5$ A
special attention was given to the cut-off of the scale free behavior
which could be associated to the finite stiffness of a testing machine
and to the hardening behavior of the material. {Recently Salman and
  Truskinovsky presented a model based on coupled Frenkel-Kontorova
  chains from which they could derive an integer-valued
  automaton\cite{Truskinovsky-PRL11}. In both versions of the model
  the dissipated energy was shown to exhibit power law avalanches with
  the same exponent $\kappa =1.6$. }

In the field of amorphous plasticity, most numerical results were
obtained using atomistic methods. Recently, in the framework of
deformation of two-dimensional Lennard-Jones model glasses, Maloney
and Robbins~\cite{Maloney-JPCM08,Maloney-PRL09} obtained a linear
dependence of the mean avalanche size with system size. Lemaitre,
Caroli and Chattoraj looked at the rate and termal dependence of the
avalanches distribution~\cite{Lemaitre-PRL09,Caroli-PRL10} and showed
that the athermal avalanche dynamics remain essentially unperturbed.

In a kinetic Monte Carlo study at mesoscopic scale, Homer {\it et al}
\cite{Homer-PRB10} identified different (stress and temperature
dependent) correlation behaviors of shear transformation zones leading
either to an avalanche-like behavior or to an homogeneous flow.  Still
at mesoscopic scale, apart from the earlier version of the present
model which considered antiplane geometry, mean fields
models~\cite{Dahmen-PRL09} have been developed by Ben-Zion, Dahmen and
collaborators in the following of a model designed by Ben-Zion and
Rice to capture earthquakes dynamics~\cite{BZR-JGR93}. Again this
class of models is very close to the one presented here with a
significant difference concerning the elastic interaction which is
assumed to be mean field.  These models predict a universal scale free
avalanche distribution with a power law exponent $\kappa =1.5$.

\begin{figure}[tbp]
\subfigure[]{
  \begin{minipage}[h]{0.48\textwidth}
   \centering
   \includegraphics[width=72mm]{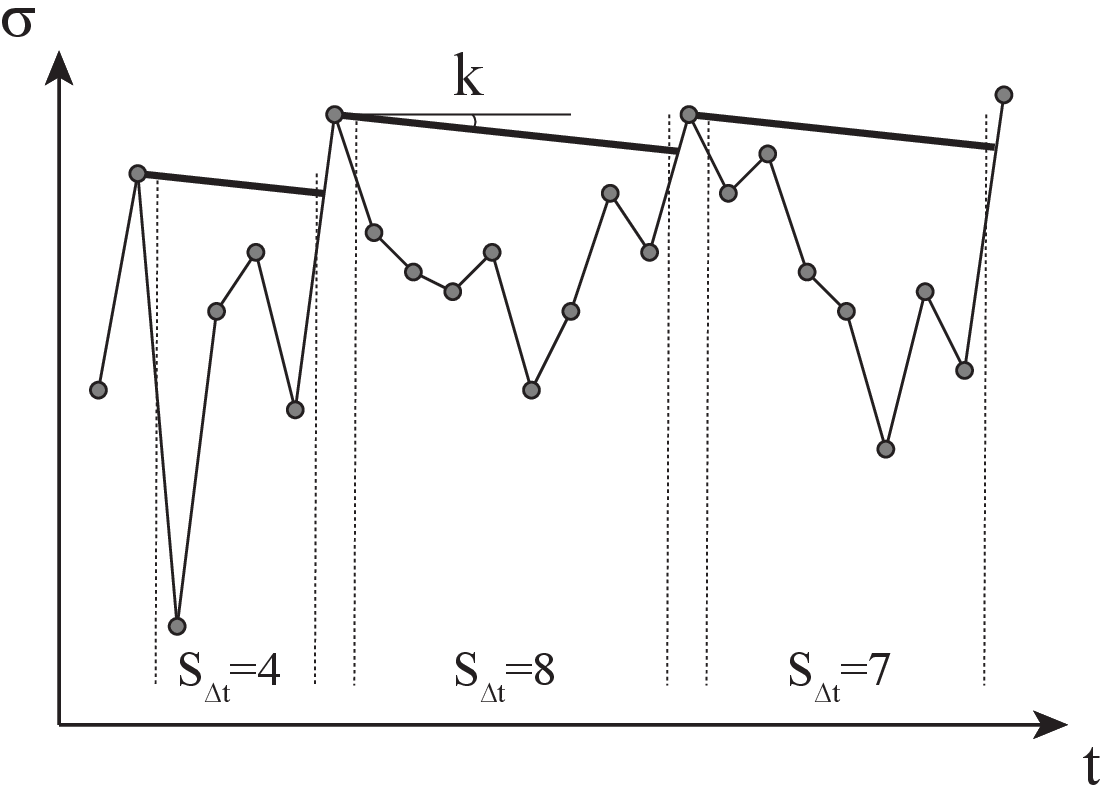}
  \end{minipage}
  \label{AvalancheDef}}
\subfigure[]{
  \begin{minipage}[h]{0.48\textwidth}
   \centering
   \includegraphics[width=76mm]{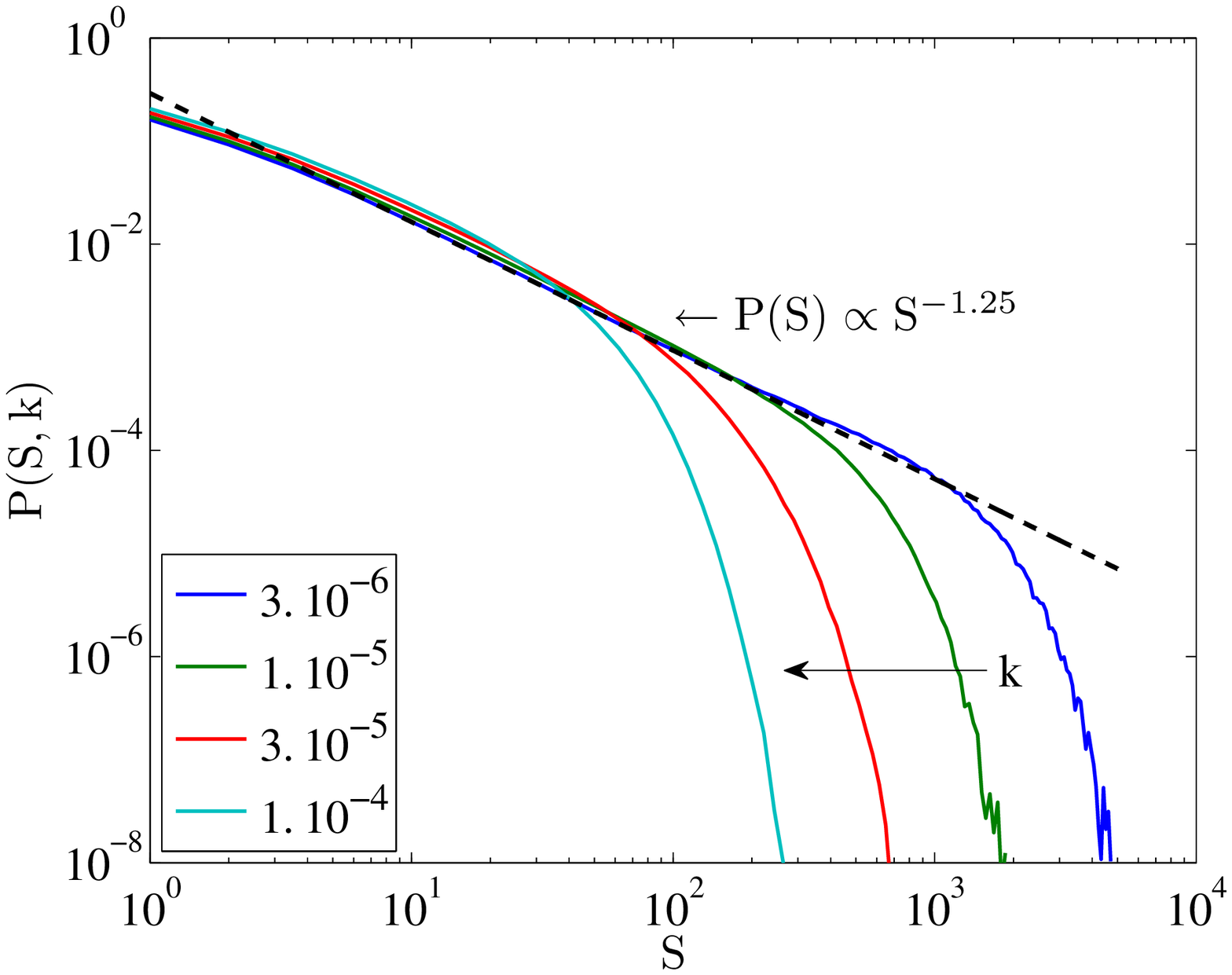}
  \end{minipage}
  \label{AvaltsLR}}
% \subfigure[]{
%  \begin{minipage}[h]{0.34\textwidth}
%   \centering
%%   \includegraphics[width=60mm]{scaling-aval.eps}
%   \includegraphics[width=57.5mm]{\FigAval/aval256-rescaled.eps}
%  \end{minipage}
%  \label{AvaltsSize1}}
  \caption{(Color online) (a) Sketch of the current yield stress (as
    obtained in extremal dynamics) (symbol $\circ$) and of the
    external stress when the system is coupled to a spring of constant
    $k$. Avalanches are defined as the intervals where the external
    stress remains larger than the yield stress.  A new avalanche is
    initiated at the next maximum yield stress after arrest. (b)
    Distributions of avalanche sizes $P(S,k)$ for a system of size
    $L=256$ and (from right to left) different values of spring
    constants $k=3. 10^{-6}, 10^{-5}, 3. 10^{-5}, 10^{-4}$. A power-law
    behavior $P(S,k)\propto S^{-\kappa}$ of exponent $\kappa\approx
    1.25$ (dashed line) is observed with a cut-off increasing as the
    spring constant decreases.
    }
  \label{AvaltsKS}
\end{figure}

\subsection{Definition and scale free behavior}

While avalanches are rather easily defined experimentally or in real-dynamics
simulations, they need to be reconstructed from the fluctuating force signal in
the case of depinning models driven through the extremal dynamics
rules~\cite{Paczuski-PRE96,Tanguy-PRE98}. Following  Ref.~\cite{Tanguy-PRE98},
avalanches are defined by introducing a small but nonzero stiffness $k$ in the
external driving as illustrated in Fig.~\ref{AvaltsKS}(a) where the bold line of
slope $-k$ represents the external driving stress. With the increasing plastic
strain the external stress is decreased linearly by a quantity $k \Delta t$,
where $\Delta t$ is the number of iteration steps from the avalanche initiation.
As soon as the driving stress drops below the current critical value $\sigma_c$,
the avalanche stops. The external spring is then loaded up to $\sigma_c$ and
trigs a new avalanche. Far from being artificial, this procedure naturally
mimics the effect of the finite stiffness of an experimental testing machine, or
the elasticity of the medium surrounding the active site~\cite{Zaiser-JSM07}.
Based on the latter argument, the thermodynamic limit of a large scale
separation between that of the STZ, and that of the medium is reproduced for a
vanishing stiffness, i.e. as for an ideal stress-controlled experiment.  One may
also note that an ideal strain-controlled experiment would be obtained for an
infinite stiffness.

The present definition size of avalanches $S=\Delta t$ should not be confused
with the duration of an avalanche measured in real time. The underlying extremal
dynamics gives no information about real time scales. The size $S$ of an
avalanche is however directly related to the strain $\varepsilon_S$ experienced
by the medium through $\varepsilon_S=S\langle \eta \rangle/L^2 $ where $\langle
\eta \rangle$ is the average of the random incremental local slip $\eta$.

The avalanche size distributions $P(S,k)$ corresponding to various stiffness
values $k$ are shown in Fig.~\ref{AvaltsKS}b. We obtain a scale free behavior
over a domain bounded by a stiffness dependent cut-off (the lower the stiffness
the larger the scaling domain). Up to the cut-off size $S^*$ avalanche
distributions follow a power law of exponent $ -\kappa$ with $\kappa=1.25 \pm
0.05$ over three decades.  This excludes the mean field value $\kappa=1.5$ as
observed in mean-field models.

{The present estimate $\kappa=1.25 \pm 0.05$ is also different
  from the results obtained on the dislocation-based models by Zaiser
  et al\cite{Zaiser-JSM05,Zaiser-JSM07} and by Salman and
  Truxskinovsky\cite{Truskinovsky-PRL11} who observe larger values of
  the scaling exponent $\kappa=1.5-1.6$. The most salient difference
  between these models and the present one is their account of short
  scale interactions between dislocations, absent in the present
  model. We note however that recent
  results~\cite{Bonamy-PRL08,Laurson-PRE10} obtained in yet a
  different framework, the propagation of an interfacial crack front,
  recently show avalanches with the very same exponent $\kappa=1.25$
  as in the present model. This observation may be far more than a
  simple coincidence. Indeed, as shown above, most of the plastic
  events occur along the directions at $\pm \pi/4$ along which the
  Eshelby elastic interaction obeys the same spatial dependence in
  $1/r^2$ as the long range elastic interaction characteristic of the
  interfacial crack growth. The latter model may then be viewed as a
  ultimate one-dimensional reduction of the present model of amorphous
  plasticity}

%In Fig.~\ref{AvaltsKS}c we show for 3 different stiffness values ($k=3\times
%10^{-6}$, $10^{-5}$, and $3\times
% 10^{-5}$) the evolution of the avalanche
%distributions for a system of size of size $L=256$ after rescaling by a power
%law of exponent $\kappa=1.25$. A clear plateau can be obtained and the cut-off
%region is well described by a gaussian extinction.

\subsection{Avalanche cut-off}

\begin{figure}[b]
 \subfigure[]{
  \begin{minipage}[h]{0.45\textwidth}
   \centering
   \includegraphics[width=80mm]{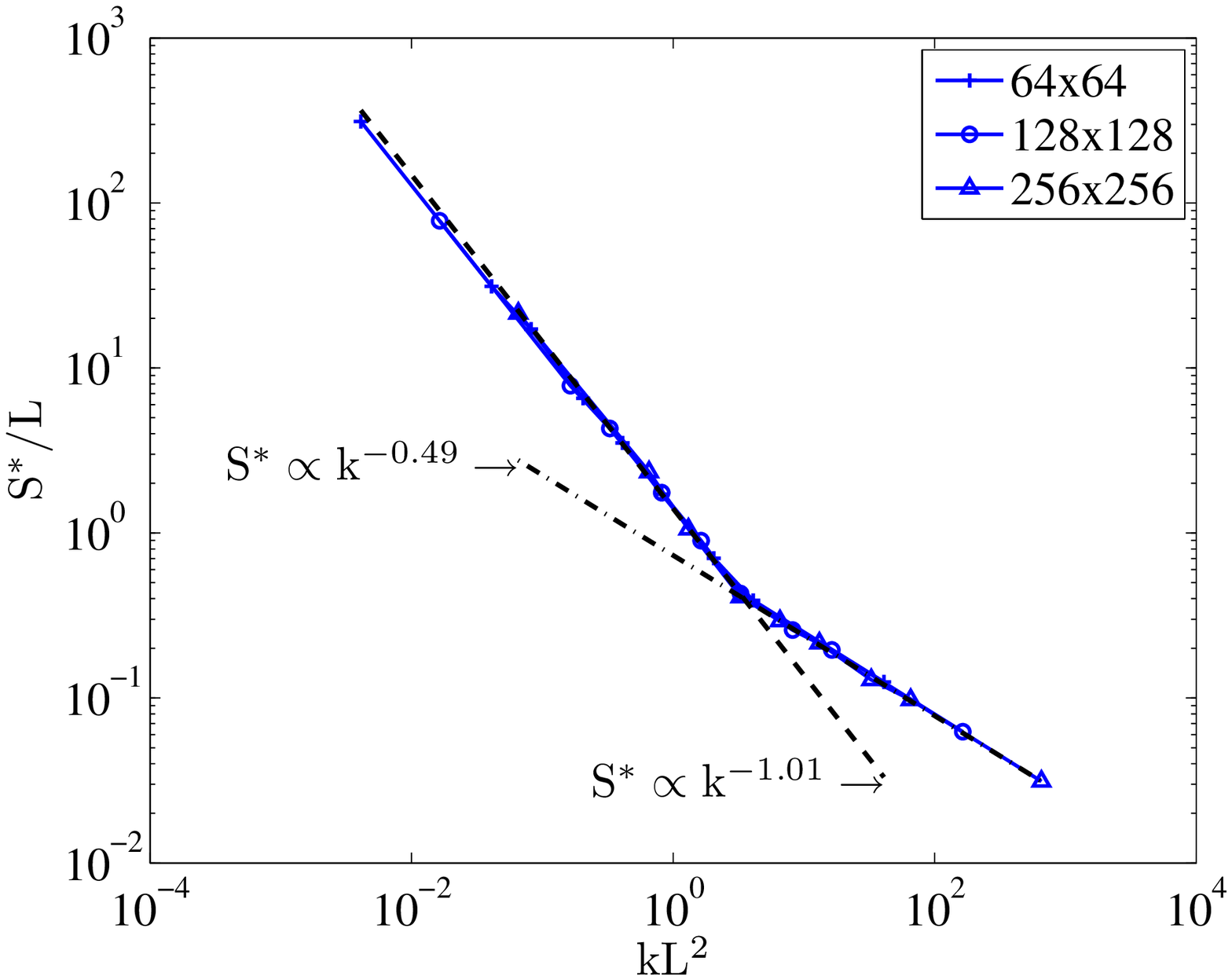}
  \end{minipage}
  \label{AvaltsSize2}}
\subfigure[]{
  \begin{minipage}[h]{0.45\textwidth}
   \centering
   \includegraphics[width=80mm]{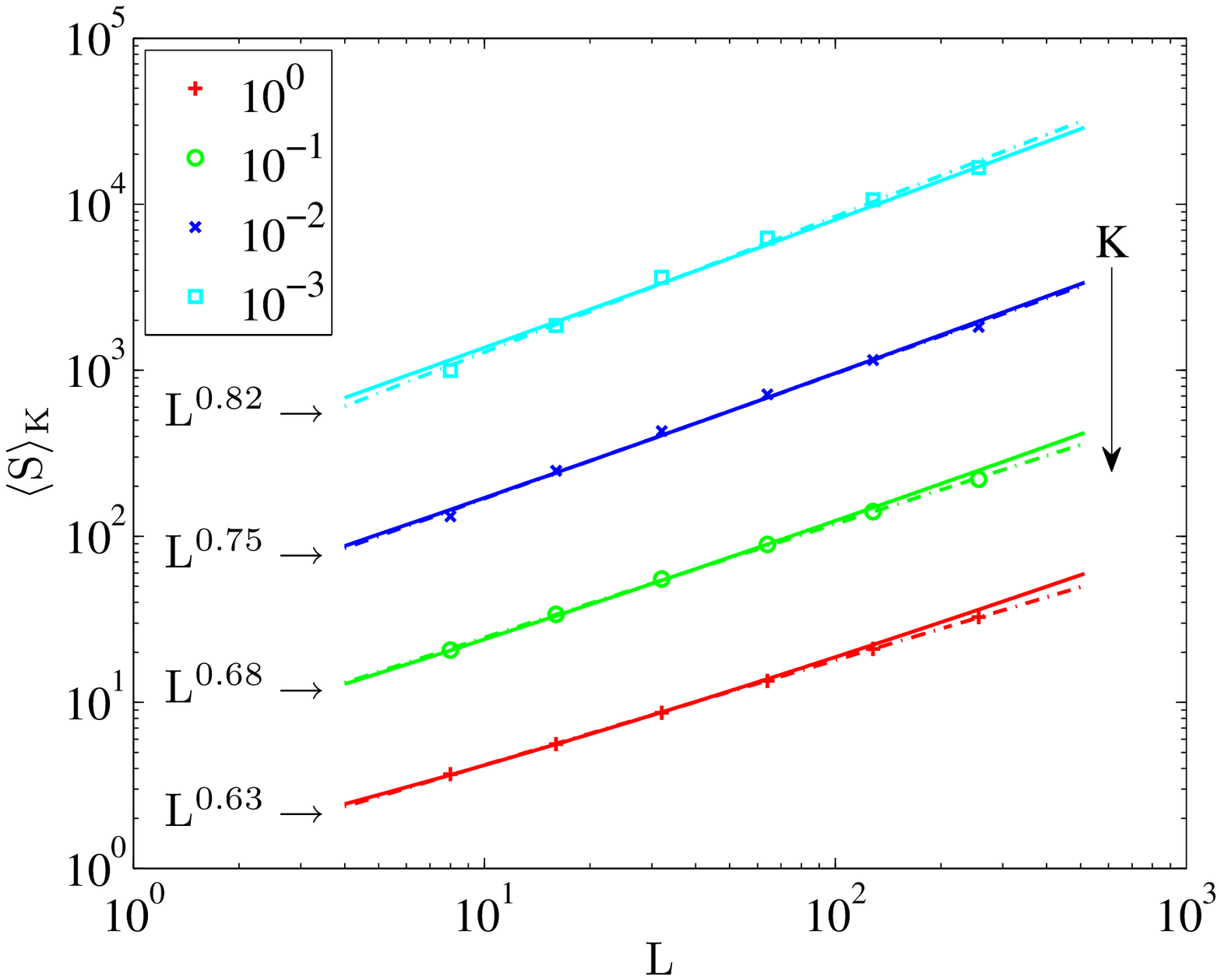}
  \end{minipage}
  \label{MStsL}}
  \caption{(Color online) (a) Scaling of the avalanche cut-off $S^*$
    {\it vs} the rescaled stiffness $K=kL^2$. The cut-off $S^*$ scales
    linearly with the system size $L$ but exhibits either an inverse
    or an inverse square root dependence on the stiffness $K$
    depending on $K$ is lower or larger than the characteristic
    stiffness $K^*=1$. (b) Mean avalanche size $\langle S
    \rangle_K$dependence on system size $L$ for stiffness values
    $K=10^{-3},\;10^{-2},\;10^{-1},\;1$. The simulation data is shown
    as symbols while the analytic expression Eq.~\ref{EQAvalSM}
    provides the continuous curves. The dash-dotted lines show the
    apparent scaling associated to these different cases. The apparent
    scaling exponents are very dependent on the stiffness value.}
  \label{AvaltsC}
\end{figure}

As discussed in ~\cite{Zaiser-JSM07} we thus could check that the introduction
of a gaussian cut-off allows us to obtain reasonable fits of the full set of
data:
\begin{equation}
P(S)\sim \Delta S^{-\kappa}\exp\left[-\left(\frac{S}{S^*}\right)^2\right]
\label{eq:powlaw}
\end{equation}
This gives us the opportunity to test the dependence of the avalanche cut-off
$S^*$ on the ``machine stiffness'' $k$.  Note here that in the framework of
extremal dynamics, obtaining a size-independent mechanical behavior (stress vs
strain) requires the change of variable $\varepsilon=t/L^{2}$ to be performed,
with $t$ being the avalanche size as above defined. Similarly, this leads to
rewrite the stiffness as $k=K/L^{2}$, $K$ being an elastic constant independent
of the system size. Looking at Fig.~\ref{AvaltsC}(a) which displays the
dependence of the avalanche cut-off size, this allows to distinguish between two
scaling regimes depending on whether the elastic constant $K$ is smaller or
larger than $K^*=1$:
\begin{equation}
S^{*} \propto
\begin{cases}
L/K \quad \text{for} \quad K < K^{*}\\
L/\sqrt{K} \quad \text{for} \quad K > K^{*}
\end{cases} \quad ,
\end{equation}

\begin{figure}[htbp]
  \subfigure[]{
  \begin{minipage}[h]{0.31\textwidth}
  \centering
  \includegraphics[width=50mm]{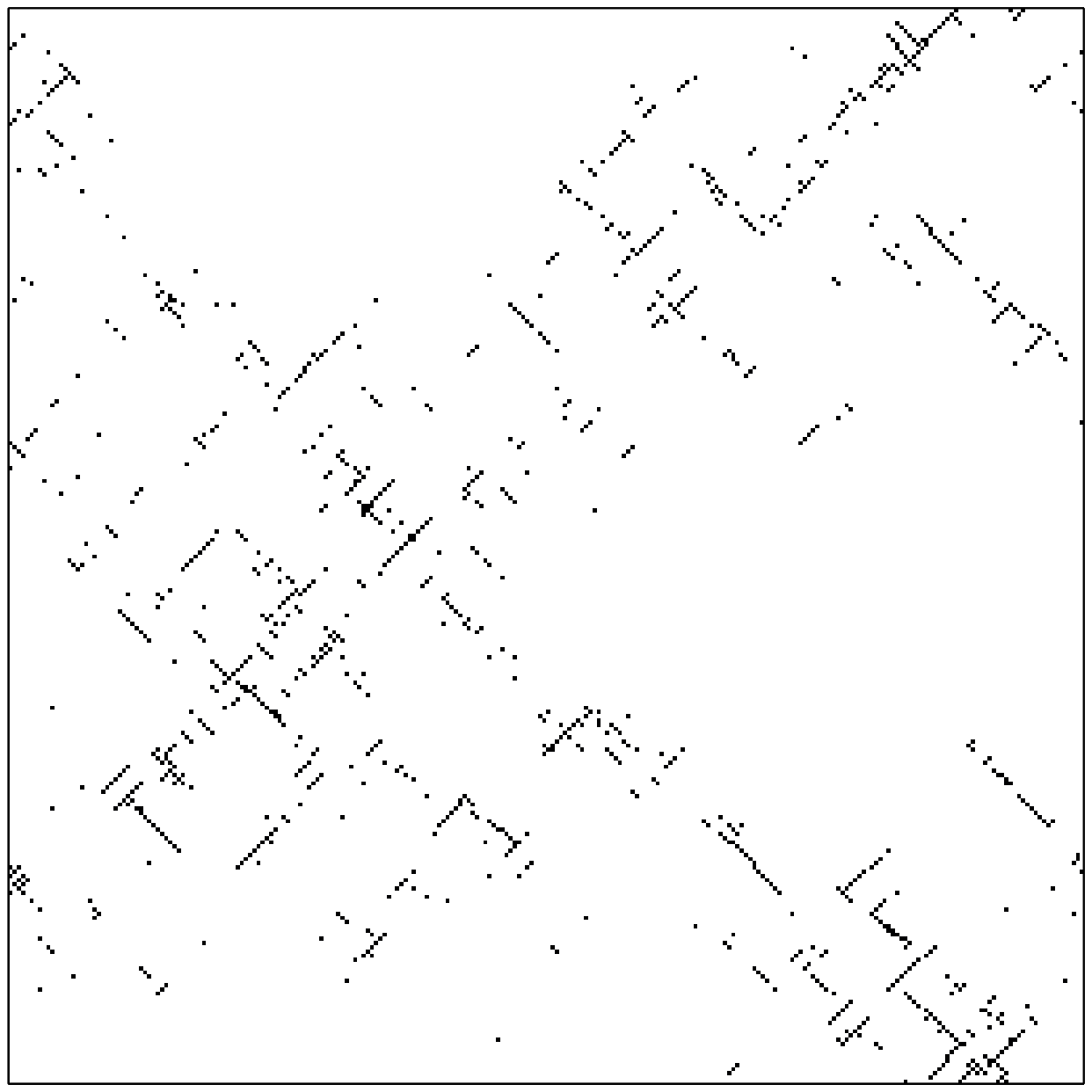}
  \end{minipage}
  \label{aval1}}
  \subfigure[]{
  \begin{minipage}[h]{0.31\textwidth}
  \centering
  \includegraphics[width=50mm]{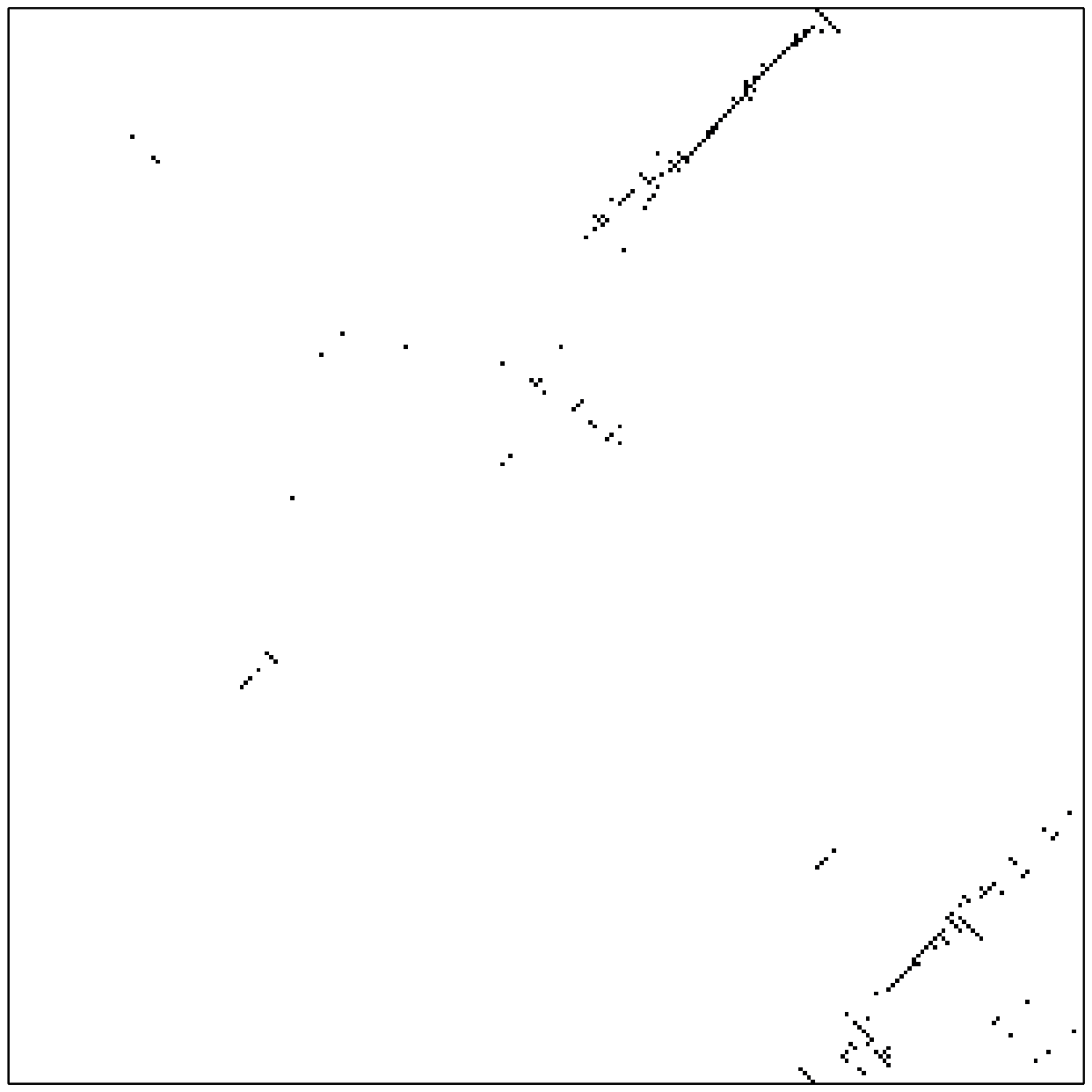}
  \end{minipage}
  \label{aval2}}
  \subfigure[]{
  \begin{minipage}[h]{0.31\textwidth}
  \centering
  \includegraphics[width=50mm]{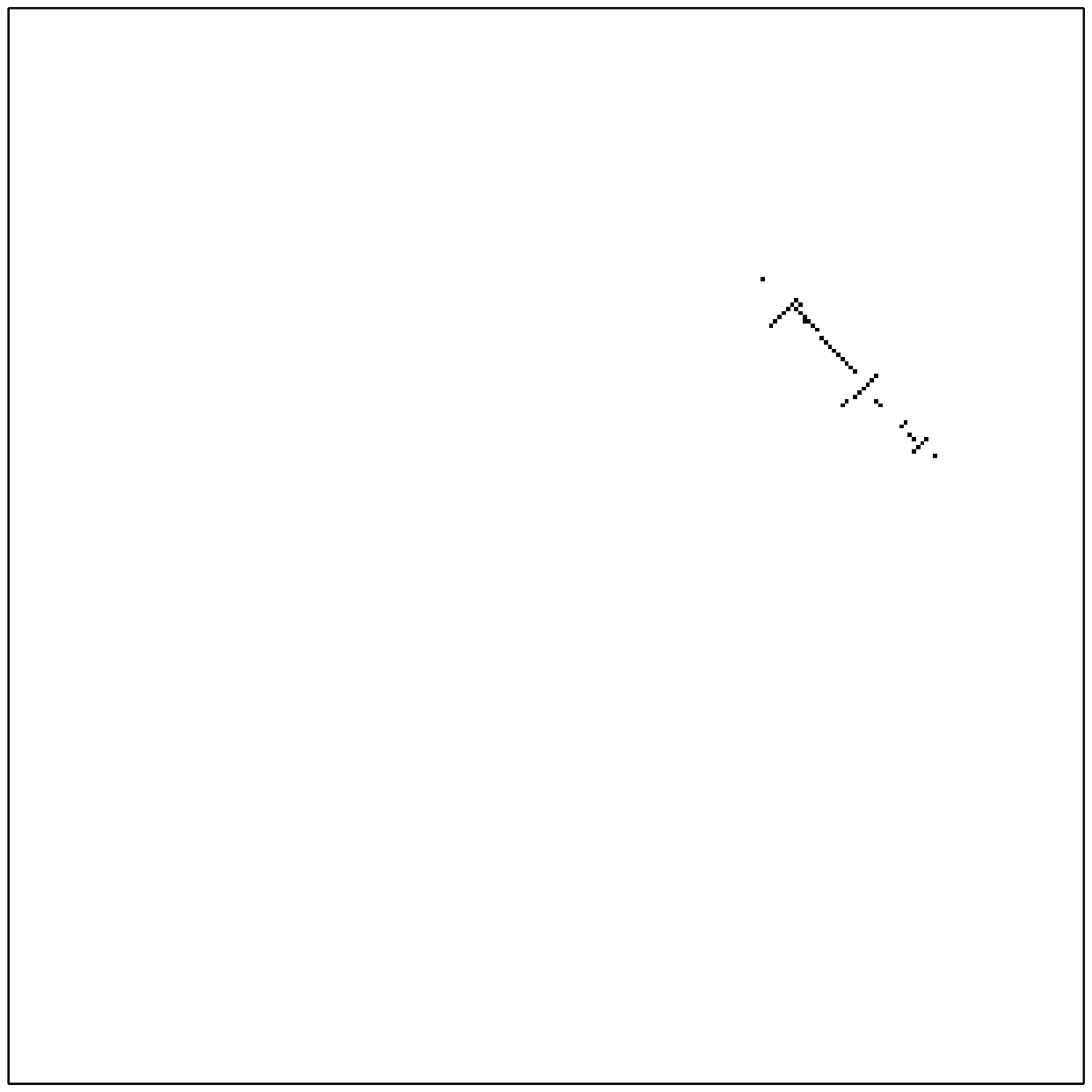}
  \end{minipage}
  \label{aval3}}
  \caption{Maps of cumulated plastic activity during avalanches obtained
  with stiffness values (from left to right) $K=10^{-1}$, $10^{0}$  and $10^{1}$
  for a system of size $L=256$. The avalanche sizes were measured to be
  $S=1616$, 324 and 81 respectively.}
  \label{AvalSCMap}
\end{figure}

In both cases we recover that the avalanche cut-off scales linearly with the
system size $L$, consistently with results by Zaiser and
Nikitas~\cite{Zaiser-JSM07}.  As illustrated in Fig. \ref{AvalSCMap}, a closer
look at the spatial structure shows that the avalanches are highly anisotropic
and once again we recover the quadrupolar symmetry of the elastic interaction.
For large values of $K$, avalanches remain mainly one-dimensional while for
lower values of $K$, two-dimensional-like patterns start to appear. One recovers
here the competition between localization (at short times) and diffusion (at
longer times) as discussed in Ref.~\cite{TPRV-Meso09a}.

\subsection{Typical size of avalanches}

With this knowledge about the avalanche cut-off the typical size  $\langle S
\rangle$ of an avalanche can thus be estimated,
\begin{equation}
\langle S \rangle  \quad \approx \frac{\int_{1}^{S^{*}} s^{1-\kappa} ds}
{\int_{1}^{S^{*}} s^{-\kappa} ds} \quad \approx \quad \frac{\kappa-1}{2-\kappa} \,
\frac{(L/K^{\upsilon})^{2-\kappa}-1}{1-(L/K^{\upsilon})^{1-\kappa}} \quad .
\label{EQAvalSM}
\end{equation}
where $\kappa\approx 1.25$, $\upsilon=1$ or $\upsilon=1/2$ depending on whether
$K<K^*$ or $K>K^*$.

In Fig.~\ref{MStsL}, we displayed the average avalanche size $\langle S \rangle$
versus the system size $L$ for several values of the elastic constant $K$.
Numerical results are well reproduced by the analytical equation
Eq.~(\ref{EQAvalSM}). Note that as soon as the elastic constants approaches
$K^*=1$, the apparent scaling can be very different from the naive scaling
obtained at very large sizes $K$ or low values of $K$: $\langle S \rangle
\propto (L/E)^{\beta}$ where $\beta=2-\kappa\approx0.75$. The value of the
machine stiffness in experimental testing is thus prone to affect the apparent
scaling of the typical size of avalanches.  Albeit the quality of the fits to
power-laws is quite satisfactory, we believe that the apparent dependence of the
exponent with $k$ is the mere reflection of the corrections which can be
rationalized by the above argument.  The present results should be compared with
those obtained by Maloney and Lema\^{\i}tre~\cite{Maloney-PRL04a} who measured an an
average avalanche size $\langle S_{\Delta t} \rangle \propto L$ on Lennard-Jones
two-dimensional model glasses under quasi-static shear and with those of
Lema\^{\i}tre and Caroli ~\cite{Lemaitre-preprint06}, $\langle S_{\Delta t} \rangle
\propto L$ or $L^{0.3}$ for a mean field model based on an effective mechanical
noise accounting for the elastic interactions.

\section{Critical threshold and Finite size fluctuations}\label{sec:finiteSize}

As discussed in the introduction, the rapid development of micro and
nano electromechanical systems is a strong motivation for the
understanding of their mechanical behavior. In such systems the ratio
between the ``micro'' size (grain domain, etc.) and the ``macro''
system size is low and strong fluctuations are expected from piece to
piece.

In that context, the present modeling of amorphous plasticity as a depinning
phenomenon if of high interest. As any phase transition, the depinning
transition exhibit finite size effects which can be quantitatively characterized
with the help of critical exponents.

In contrast to usual studies of criticality, the present context invites us to
focus on the critical threshold rather than on the critical exponents. While the
former is reputed of low interest since it depends on the microscopic details,
it gives us here the yield stress value at macroscopic scale. The fluctuations
of this threshold for finite systems will thus directly give the expected
fluctuations of the yield stress for small pieces.

We develop below a study of finite-size effects which follow the lines of
previous works about the depinning of elastic lines~\cite{VSR-PRE04}.

As defined in the description of the model, each elementary zone is
characterized by a local plastic criterion $\sigma_{c}(\bm
x)=\sigma_{\gamma}(\bm x)-\sigma_{\mathrm{res}}(\bm x)$ (where $\bm x$
refers to the spatial location) which can be separated in two
contributions: $\sigma_{\gamma}$ corresponds to the yield threshold of
the local structure in absence of internal stress;
$\sigma_{\mathrm{res}}$ is the internal stress induced by the
successive plastic reorganizations that have occurred in the
material. For each configuration of the system, a loading that do not
trigger any local slip event obeys $\Sigma^{ext} < \Sigma_c =
\min_{\bm x} \sigma_c(\bm x)$ and the macroscopic yield stress is then
defined as the maximum value of this macroscopic load over the whole
set of configurations, $\Sigma^*=\max \Sigma_c$; when the external
stress lies below that value, $\Sigma^*$, plastic deformation will
eventually stop after a finite strain while above it the material can
flow indefinitely.  To recast this definition in the previous language
of avalanches, the macroscopic yield stress is the one which
corresponds to the existence of an infinite size avalanche at
vanishing stiffness $k=0$.

\begin{figure}[htbp]
\subfigure[]{
  \begin{minipage}[h]{0.31\textwidth}
   \centering
   \includegraphics[width=55mm]{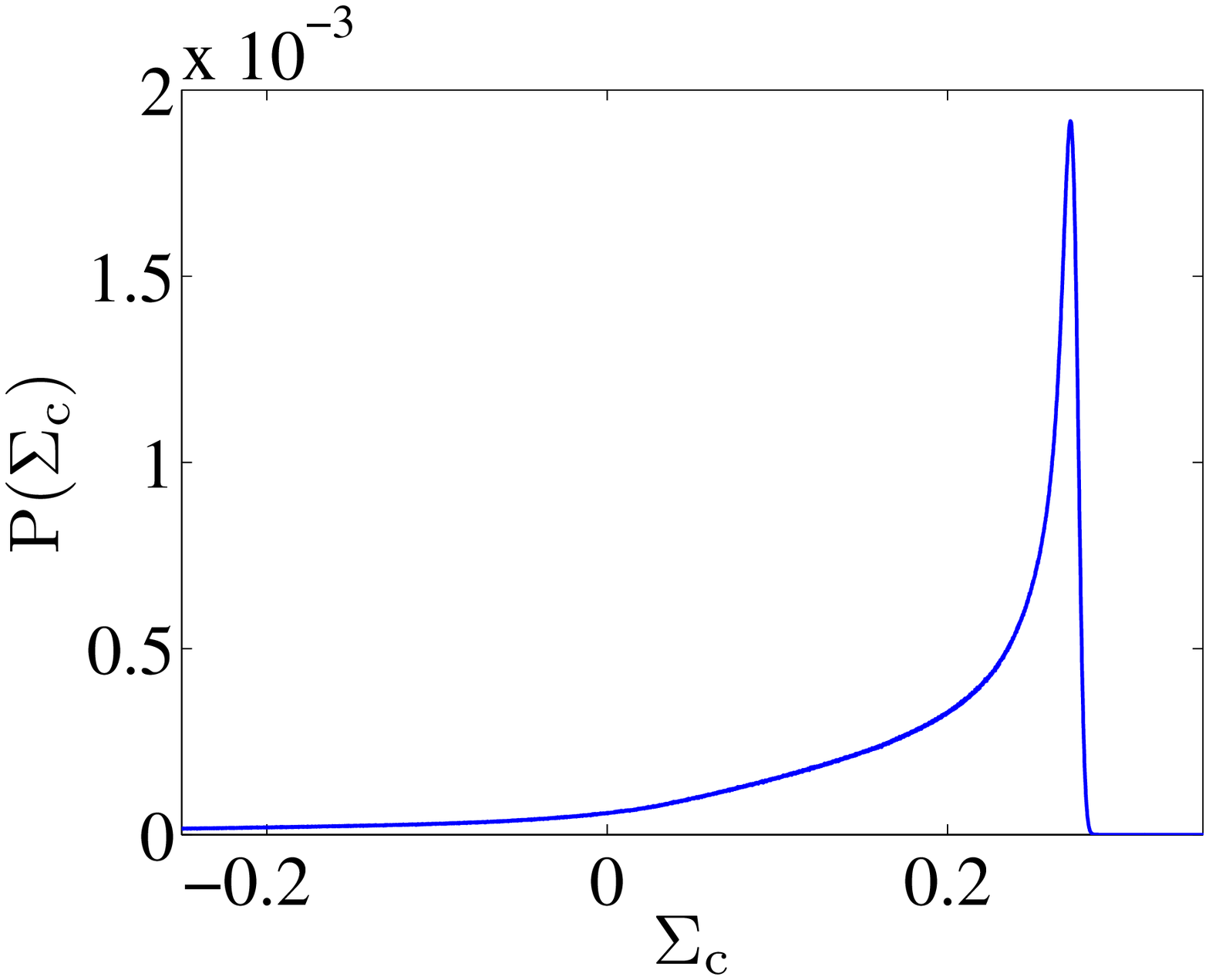}
  \end{minipage}
  \label{DistCF}}
 \subfigure[]{
  \begin{minipage}[h]{0.31\textwidth}
   \centering
   \includegraphics[width=55mm]{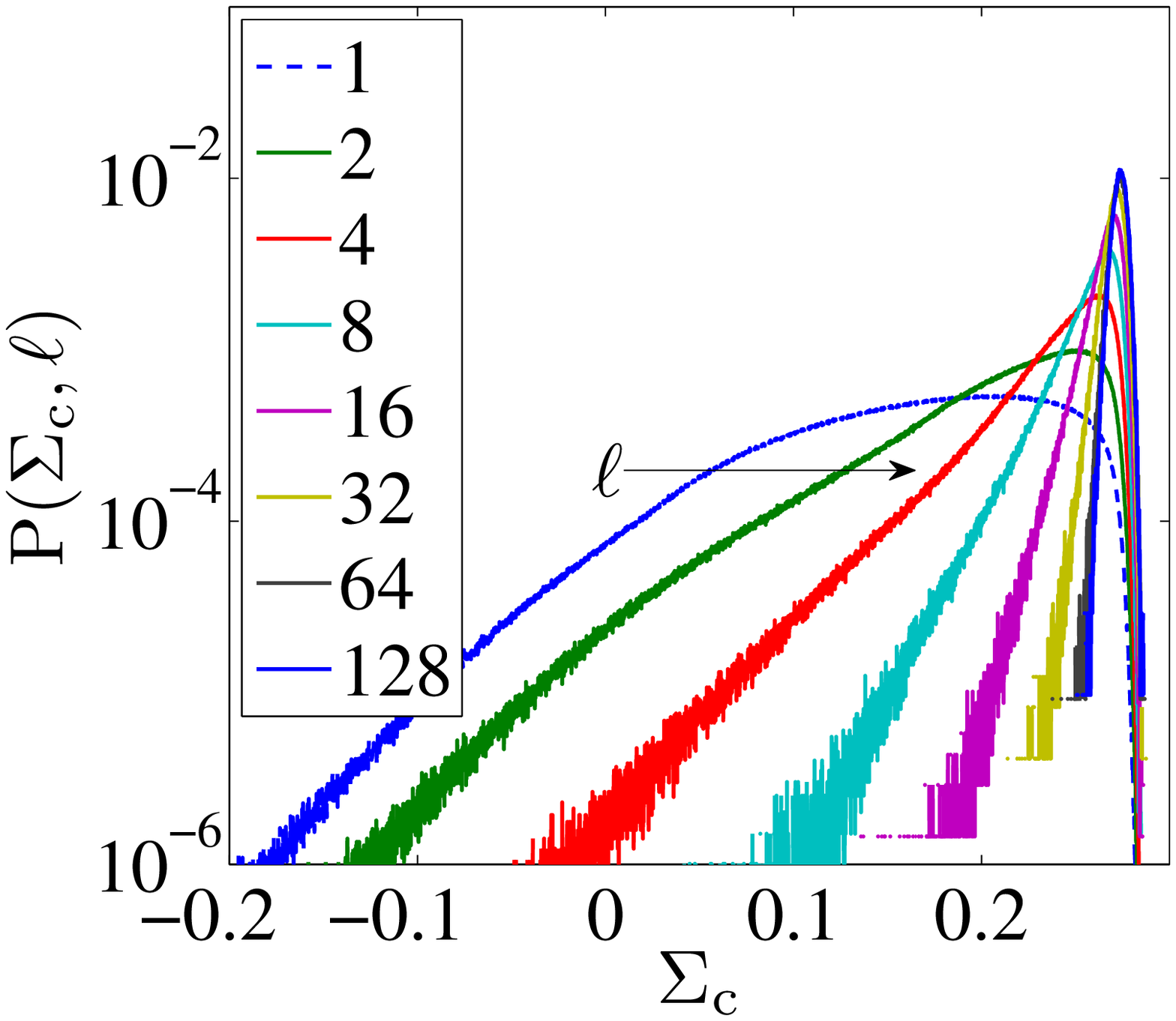}
  \end{minipage}
  \label{DistCFddL}}
 \subfigure[]{
  \begin{minipage}[h]{0.31\textwidth}
   \centering
   \includegraphics[width=55mm]{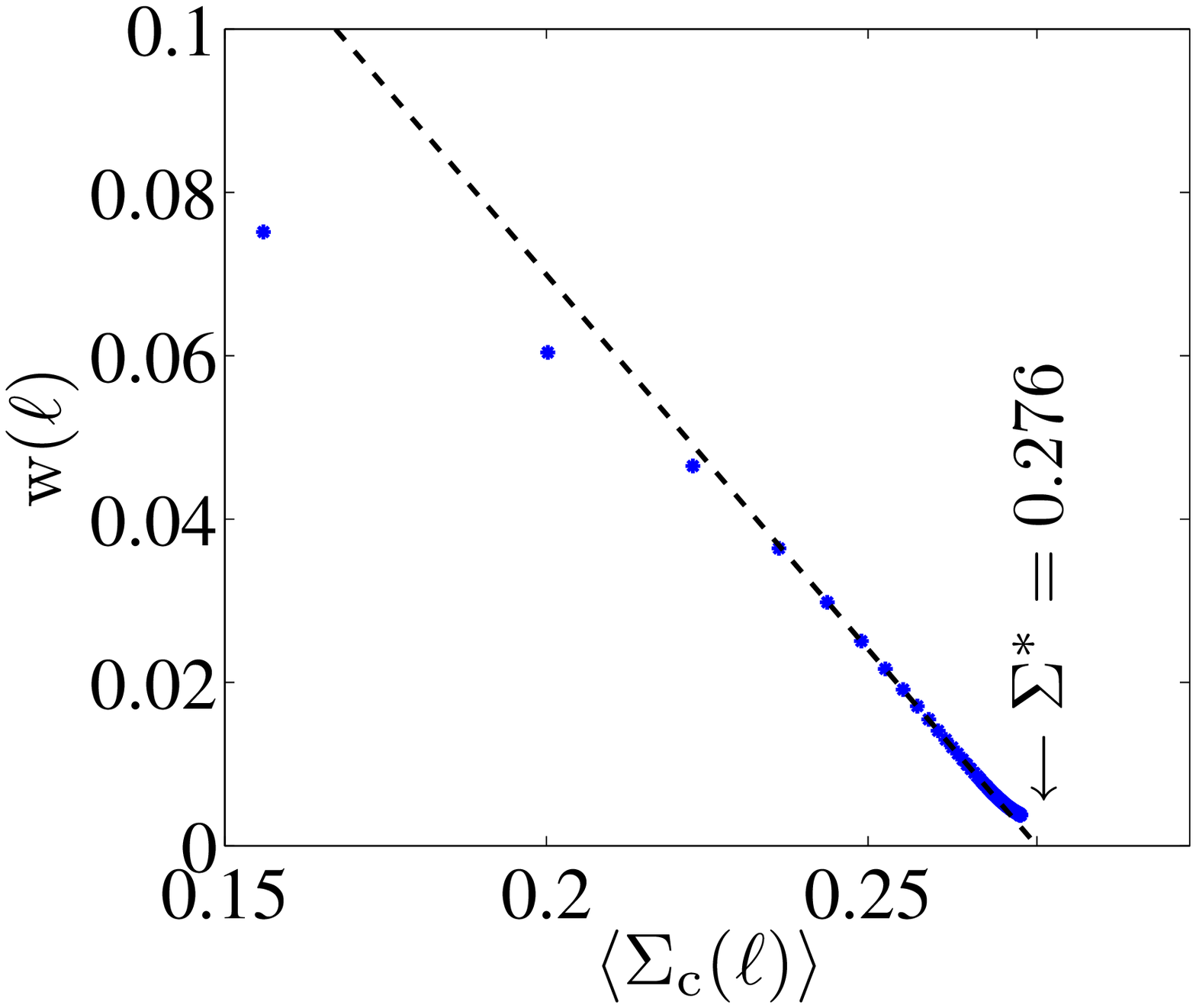}
  \end{minipage}
   \label{pcfmsd}}
\caption{(Color online) (a) Distribution $P(\Sigma_c)$ of the current
  yield stress $\Sigma_c$ sampled over time.  The macroscopic yield
  stress $\Sigma^*$ is given by the maximum of the distribution.  (b)
  Representation in semi-logarithmic coordinates of the conditional
  distributions $P(\Sigma_c,\ell)$ such that the next slip event is
  located at a distance $\ell$ from the current point.
  $P(\Sigma_c,\ell)$ can be interpreted as the distribution of yield
  stress fluctuations for a subsystem of size $\ell$.  (c) Dependence
  of the width $w$ of the distributions $P(\Sigma_c,\ell)$ {\it vs.}
  its mean value $\langle\Sigma_c (\ell)\rangle$ for different
  $\ell$. The linear behavior obtained for large values of $\ell$
  allows to estimate the asymptotic yield limit $\Sigma^*$ through
  extrapolation to $w=0$.}
  \label{LocalForces}
\end{figure}

The distribution of these current yield stress values $P(\Sigma_c)$ is
shown in Fig. \ref{LocalForces}(a). Note that $\Sigma_c$ can take here
negative or positive values since it is associated to a fluctuating
part of the material properties. (Changing the material yield limit
will trivially translate both the distribution and the critical
threshold $\Sigma^*$.) The same distribution is shown in
\ref{LocalForces}(b) in logarithmic scale.  The entire distribution
depends on the local yield stress distribution, here a uniform
distribution, and hence has no specific value.  However, close to its
maximum, the distribution contains only information relative to
macroscopically pinned or quasi pinned configurations.  Hence the
behavior of the distribution close to the maximum stress contains
generic features which are difficult to extract from such a graph.

To isolate those universal features, it is proposed to part the
distribution into conditional probability distribution functions
depending on a characteristic which signals that the configuration is
close to pinning. We chose the distance $\ell$ between successive slip
events as a clear indication of such a pinned configuration.  $P(
\Sigma_c,\ell)$ is introduced as the fraction of the initial
distribution $P( \Sigma_c)$ such that the plastic event occurred at a
distance $\ell$ from the previous one. This trick gives us a simple
way of analyzing finite size effects. Indeed, writing that plastic
activity has to move by a distance $\ell$ simply means that over a
domain of extension $\ell$, the system has reached a pinned
configuration. $P( \Sigma_c,\ell)$ thus gives direct access to the
distribution of effective thresholds for systems of size $\ell$. We
observe that the larger the distance (the system size), $\ell$ , the
narrower the distribution $P( \Sigma_c,\ell)$ and the closer its
center from the maximum of the distribution $P( \Sigma_c)$.  This
observation is rationalized in Fig. \ref{LocalForces}(c) where the
width of these conditional distributions is plotted against their
mean. We obtain a linear behavior,{\it i.e.,} these two quantities
obey the same scaling. In particular, this means that extrapolating
this linear behavior to a zero width (which would be obtained for an
infinite system) allows us to give a precise estimate of the critical
threshold $\Sigma^*$.

\begin{figure}[htbp]
 \subfigure[]{
  \label{fig:mini:subfig:a}
  \begin{minipage}[h]{0.45\textwidth}
   \centering
   \includegraphics[width=80mm]{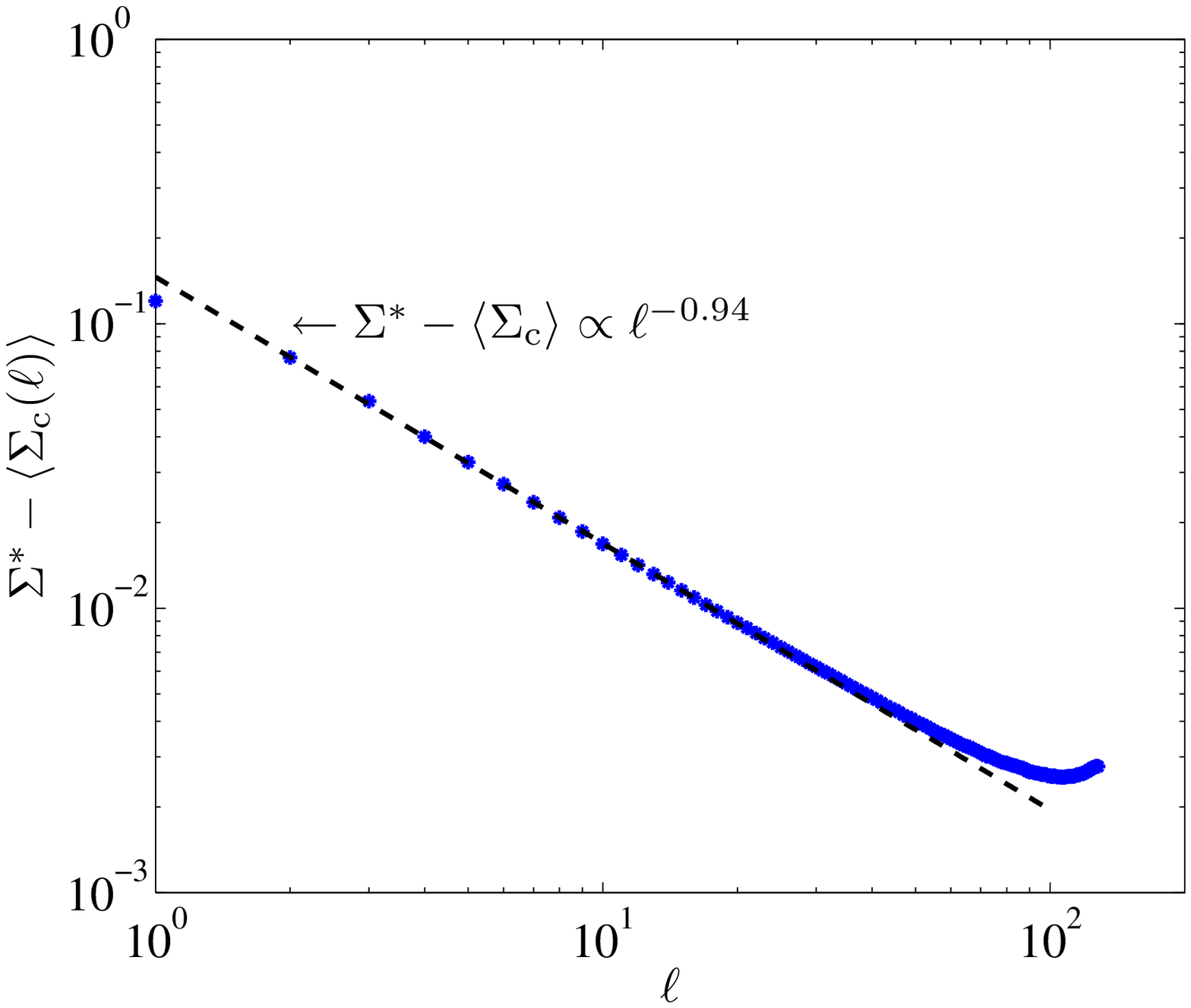}
  \end{minipage}
  \label{PCFM}}
 \subfigure[]{
  \label{fig:mini:subfig:b}
  \begin{minipage}[h]{0.45\textwidth}
   \centering
    \includegraphics[width=80mm]{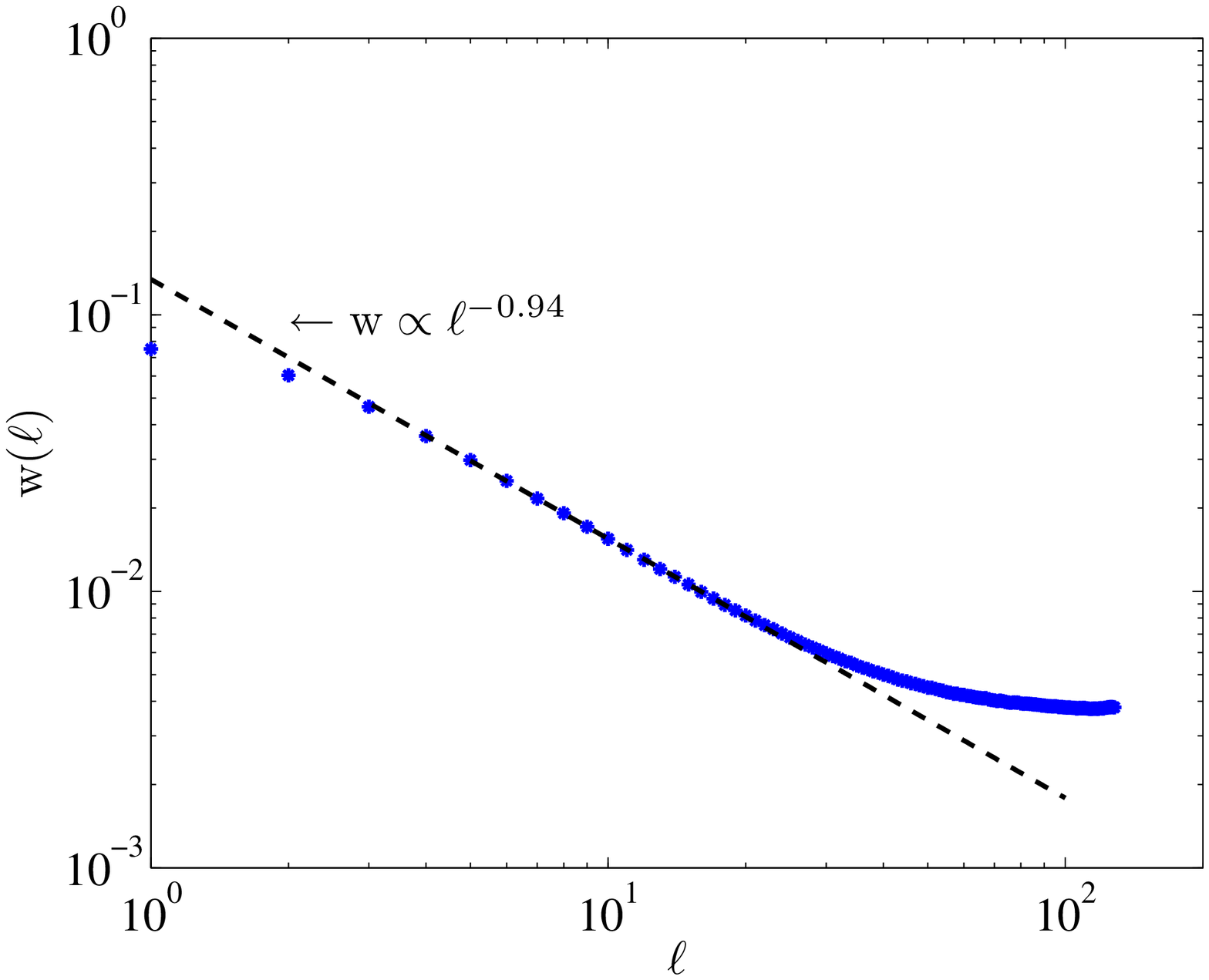}
 \end{minipage}
  \label{PCFSD}}
\caption{(Color online) (a) shows the difference between the
  asymptotic macroscopic yield stress, $\Sigma^*$ and the mean yield
  stress $\langle\Sigma_c(\ell)\rangle$ {\it vs.} distance $\ell$, for
  a system of size $L=256$.  A power-law fit of exponent
  $b\approx0.94$ is shown as a dotted line.  (b) is a plot of the
  standard deviation of depinning stress $w$ as a function of the
  distance $\ell$, for a system of size $L=256$.  A similar power-law
  fit of exponent $b\approx0.94$ is shown as a dotted line.  }
  \label{PCFMnSD}
\end{figure}

The precise knowledge of the critical threshold gives us the opportunity to
characterize not only the scaling behavior of the finite size fluctuations of
the yield stress but also of the distance between the mean yield stress and the
critical threshold. The two scaling behaviors are displayed in
Fig.~\ref{PCFMnSD}. We see that for a typical size $\ell$ both the width
$w(\ell)$ of the yield stress fluctuations and the distance to threshold
$\Sigma^*-\Sigma_c(\ell)$ obey the same scaling:
    \be
    w(\ell) \propto \ell^{-b}\; ;
    \quad \Sigma^*-\Sigma_c(\ell) \propto \ell^{-b}
    \quad \textrm{with}
    \quad b\approx 0.94
    \ee
The macroscopic yield stress is hampered by finite size systematic corrections
roughly inversely proportional to the system size.  While in the context of
elastic line depinning, a similar power-law correction was observed, the
exponent $b$ could be related to the roughness exponent~\cite{VSR-PRE04}, in the
present case, we could not build a similar scaling relation.

\section{Plastic precursors}\label{sec:precursor}

As discussed in Ref.~\cite{TPRV-Meso09a}, the plastic strain obeys a strong
anisotropic scaling resulting from the quadrupolar symmetry of the elastic
stress redistribution. Most of the plastic events occur consecutively to a
previous plastic event located along a direction at $\pm \pi/4$ (maximum shear
directions).

\begin{figure}[htbp]
\subfigure[]{
  \begin{minipage}[h]{0.45\textwidth}
   \centering
   \includegraphics[width=80mm]{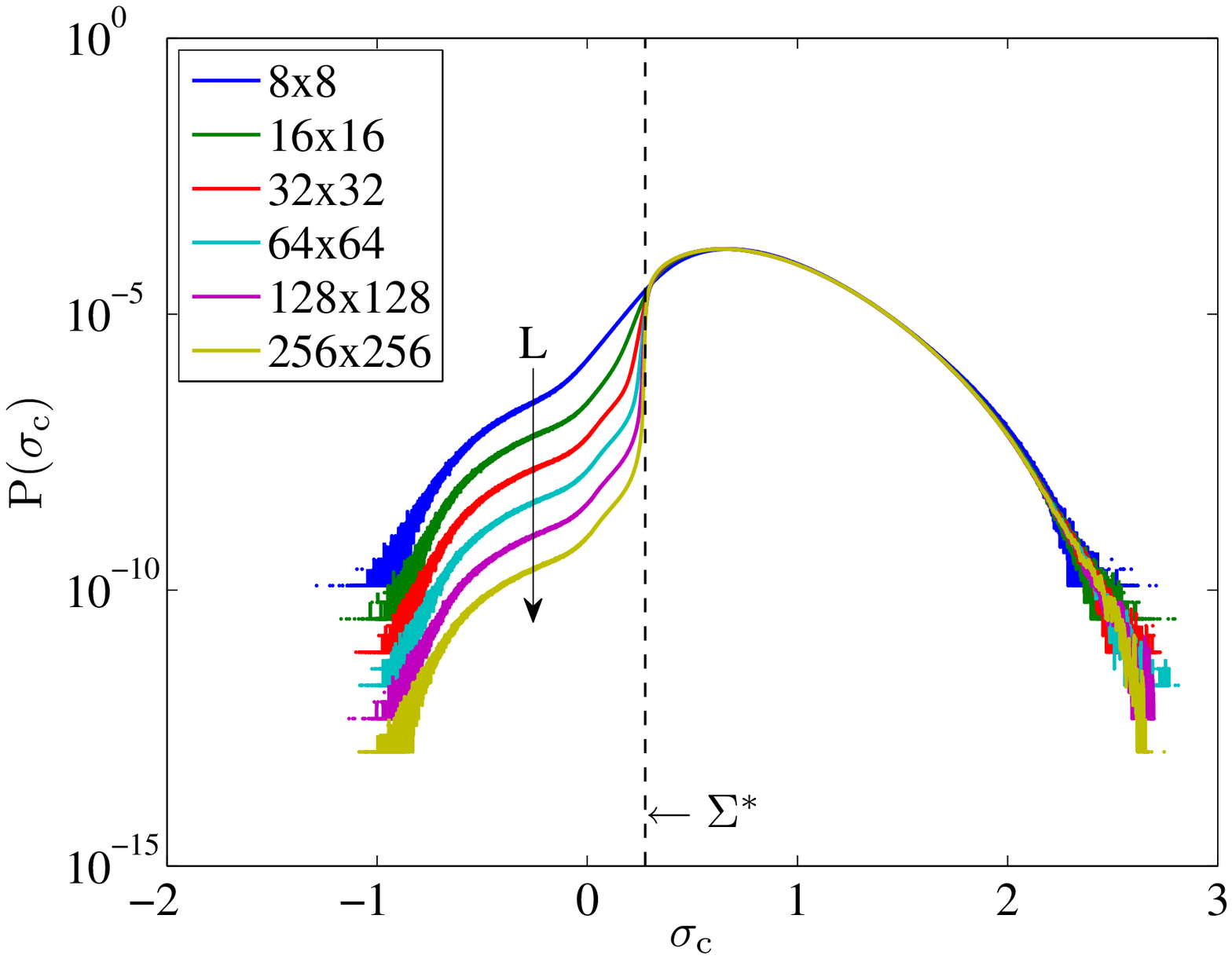}
  \end{minipage}
  \label{DistLF}}
 \subfigure[]{
  \begin{minipage}[h]{0.45\textwidth}
   \centering
   \includegraphics[width=80mm]{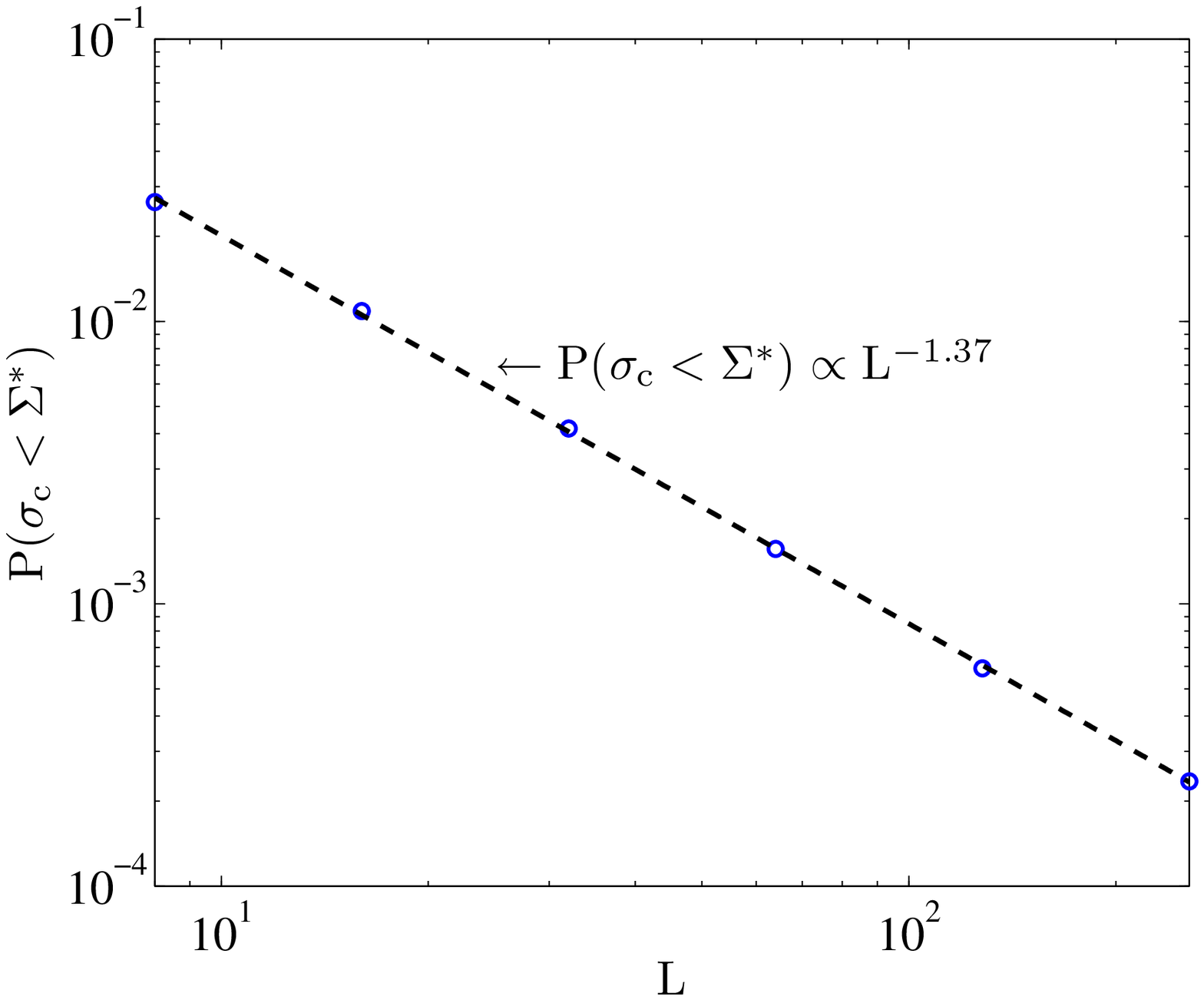}
  \end{minipage}
  \label{CDistLFL}}
  \caption{(Color online) (a) Distribution of individual depinning
    stress $P(\sigma_{c}(\bm x))$ for different system sizes $L=8$,
    16, 32, 64, 128 and 256; The dotted line indicates the macroscopic
    yield threshold $\Sigma^{*}$: The overcritical part
    ($\sigma_{c}<\Sigma^{*}$) depends on $L$ (in contrast to the
    subcritical one).  (b) The relative weight of the overcritical
    part $P(\sigma_{c}<\Sigma^{*})$ is observed to scale as a
    power-law of the system size.  $P(\sigma_{c}<\Sigma^{*})\propto
    L^{-s}$ with $s\approx1.34$.}
  \label{subcritical}
\end{figure}

However, it appears that only a tiny fraction of the sites is prone to slip. It
is possible to distinguish these precursory sites when looking at the full
distribution of the local plastic thresholds $\sigma_c(\bm x)$. Such
distributions are displayed in Fig.~\ref{subcritical}(a) for different system
sizes. The dashed vertical line at the abscissa of the critical threshold
$\sigma^*$ allows us to separate two populations. The left part corresponds to
the weakest sites of the lattice. For one particular configuration, the weakest
site gives the current yield stress $\Sigma_c$. The right part corresponds to
the sub-critical sites, their local threshold being larger than the critical
value $\Sigma^*$ they are unconditionally stable. For one configuration however,
not only the weakest site but also a few others can be characterized by a
overcritical local threshold {\it i.e.} $\sigma_c(\bm x) < \Sigma^*$. They are
thus very likely to initiate a slip event, and hence can be termed
``precursors''. A scaling analysis of this population is of interest. It is
clear from Fig.~\ref{subcritical}(a) that the fraction of precursors decreases
when the size $L$ of the system increases. The right panel \ref{subcritical}(b)
shows the size dependence of this population in logarithmic scale. We obtain
$P[\sigma_c(\bm x) < \Sigma^*] \propto L^{-s}$ with $s\approx 1.34$. This
scaling can be interpreted as the fact that these precursory sites live on a
fractal support of dimension $d_F = 2-s \approx 0.66$.

\begin{figure}[htbp]
  \centering
   \includegraphics[width=57mm]{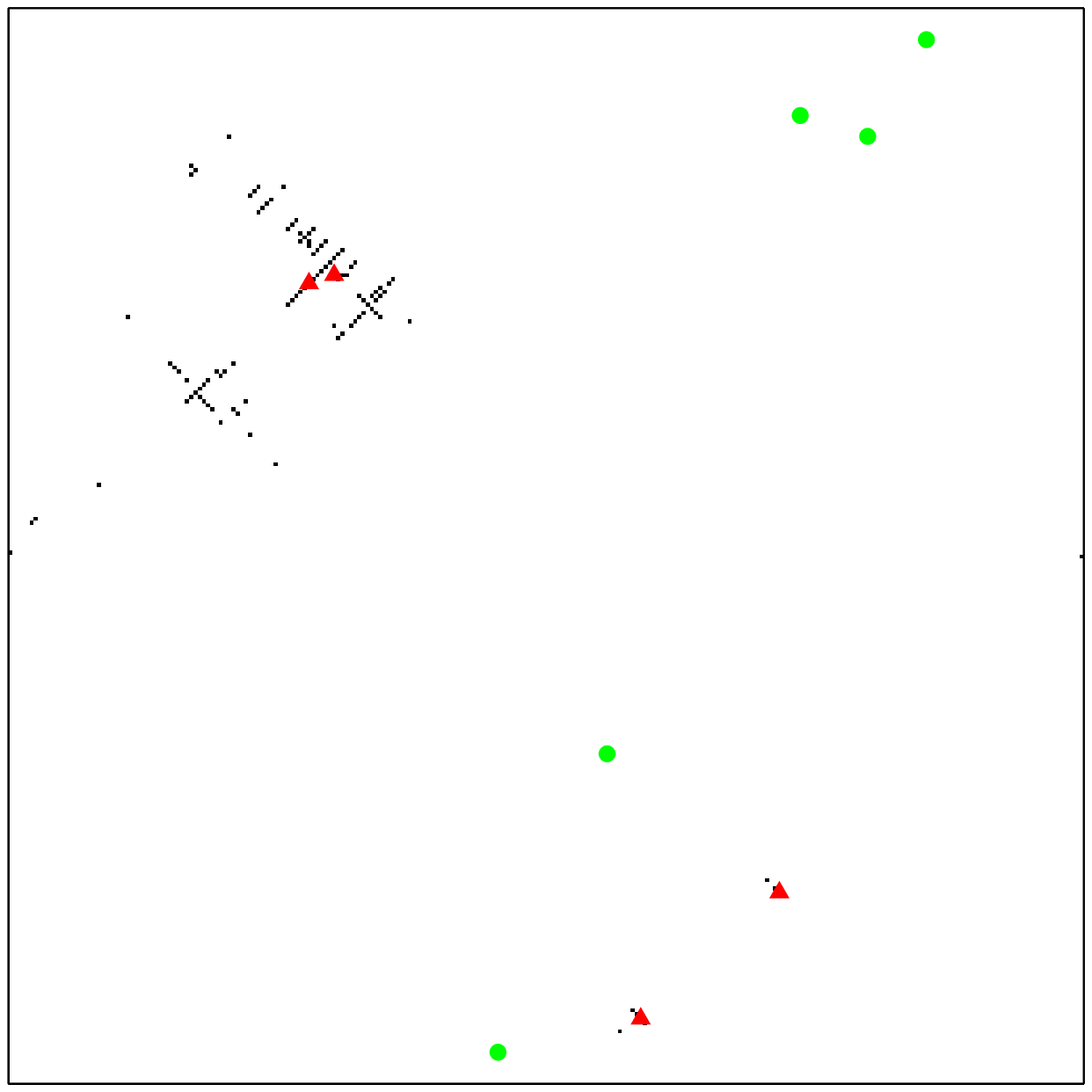}
   \includegraphics[width=57mm]{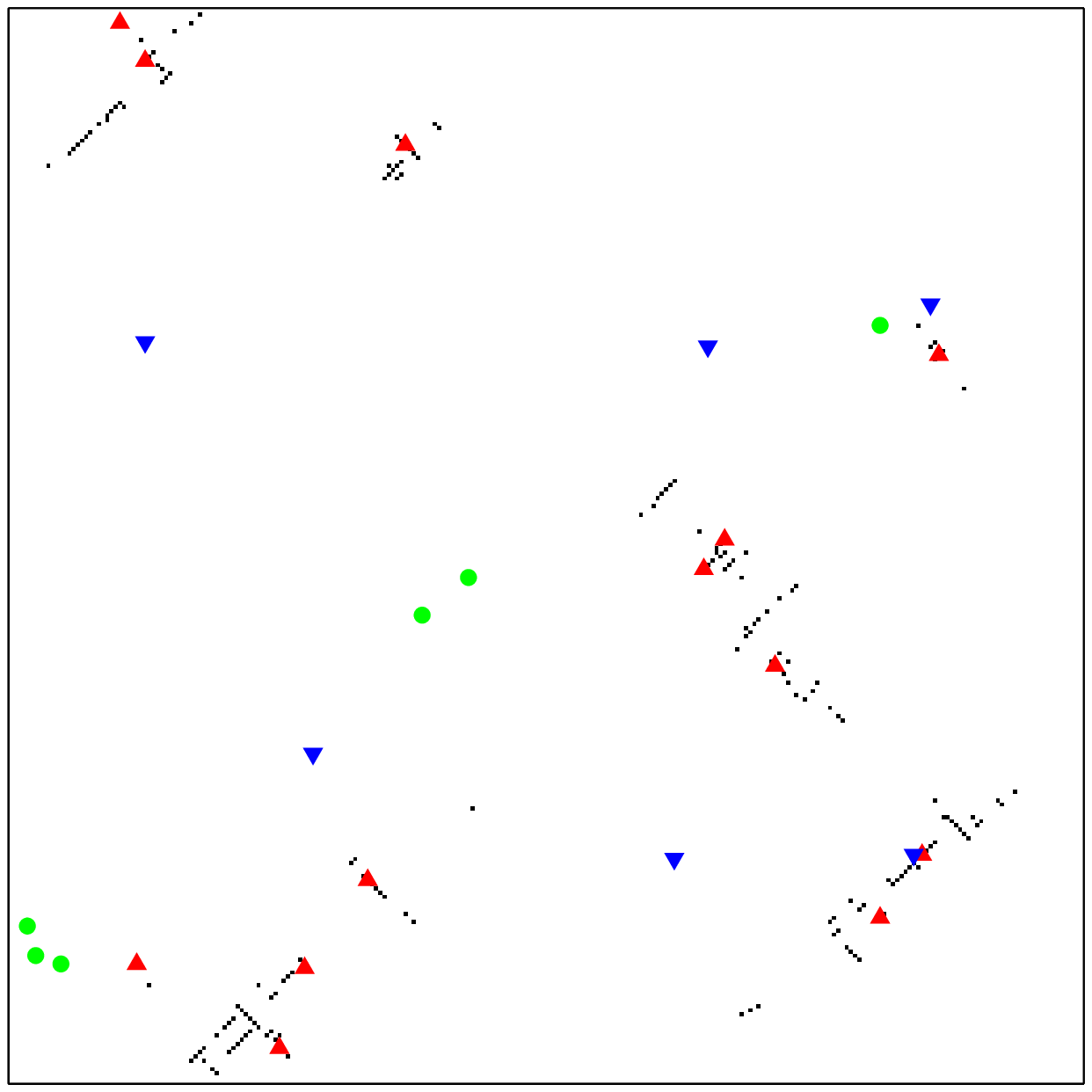}
   \includegraphics[width=57mm]{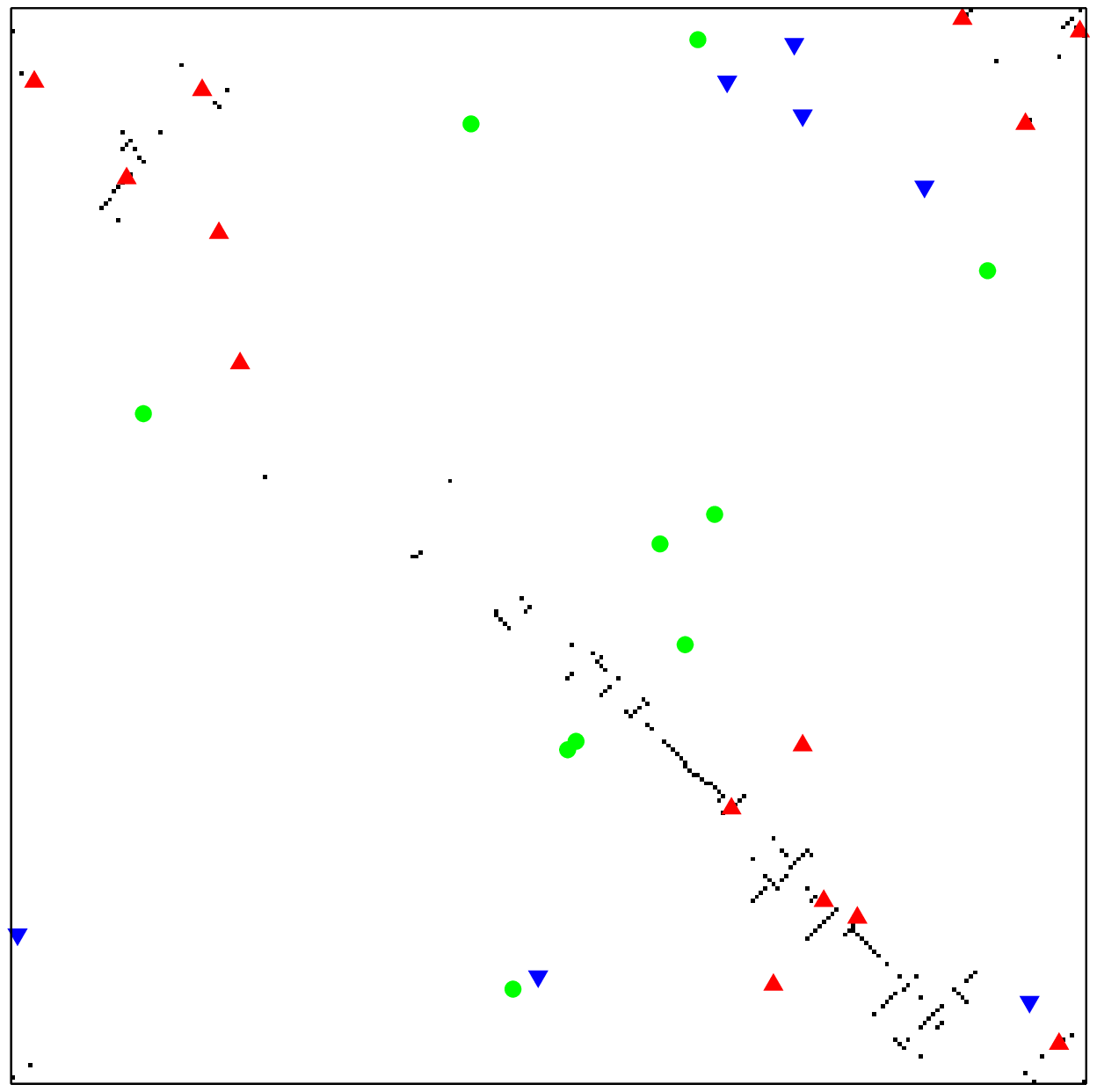}
   \includegraphics[width=57mm]{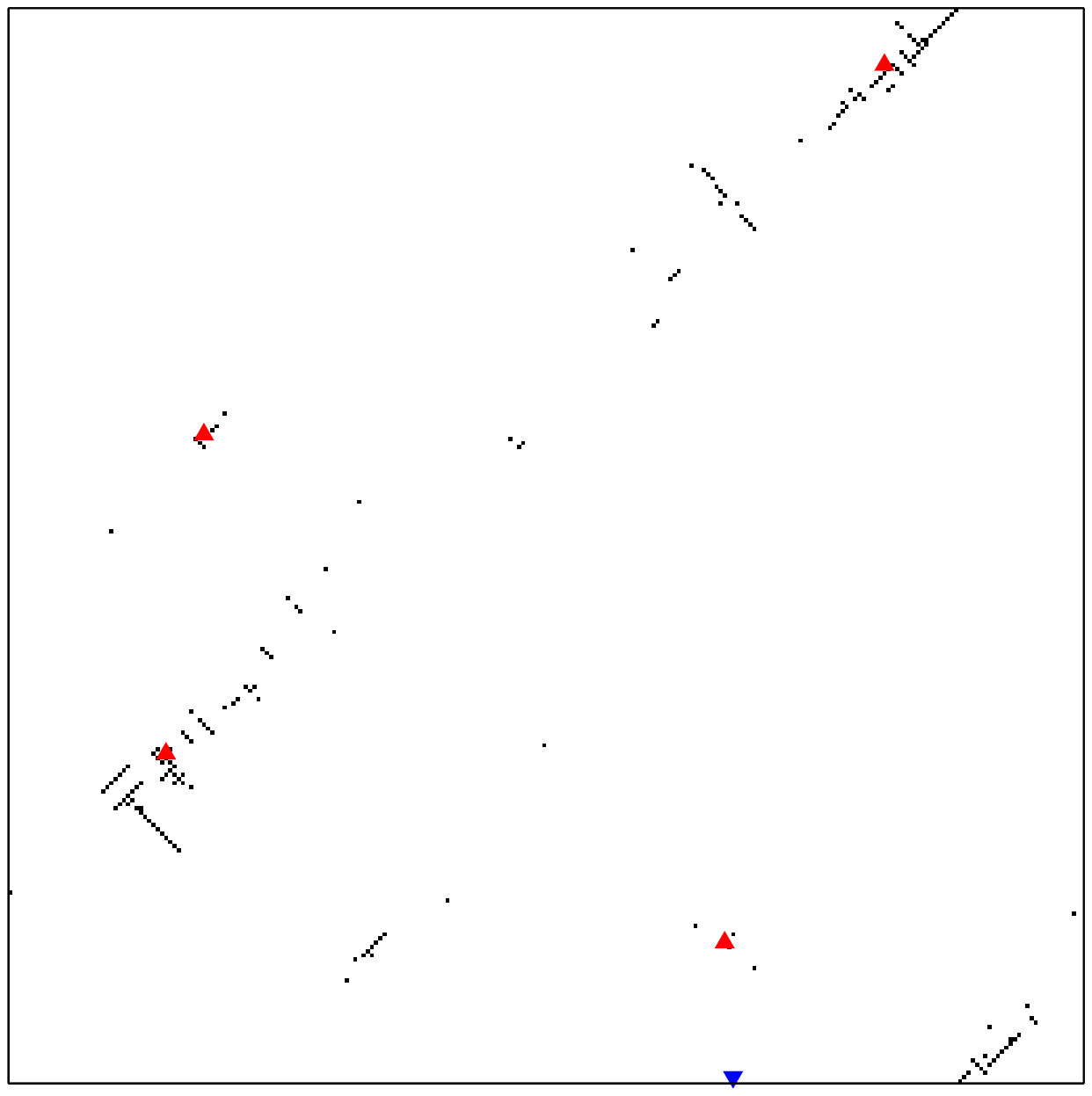}
   \includegraphics[width=57mm]{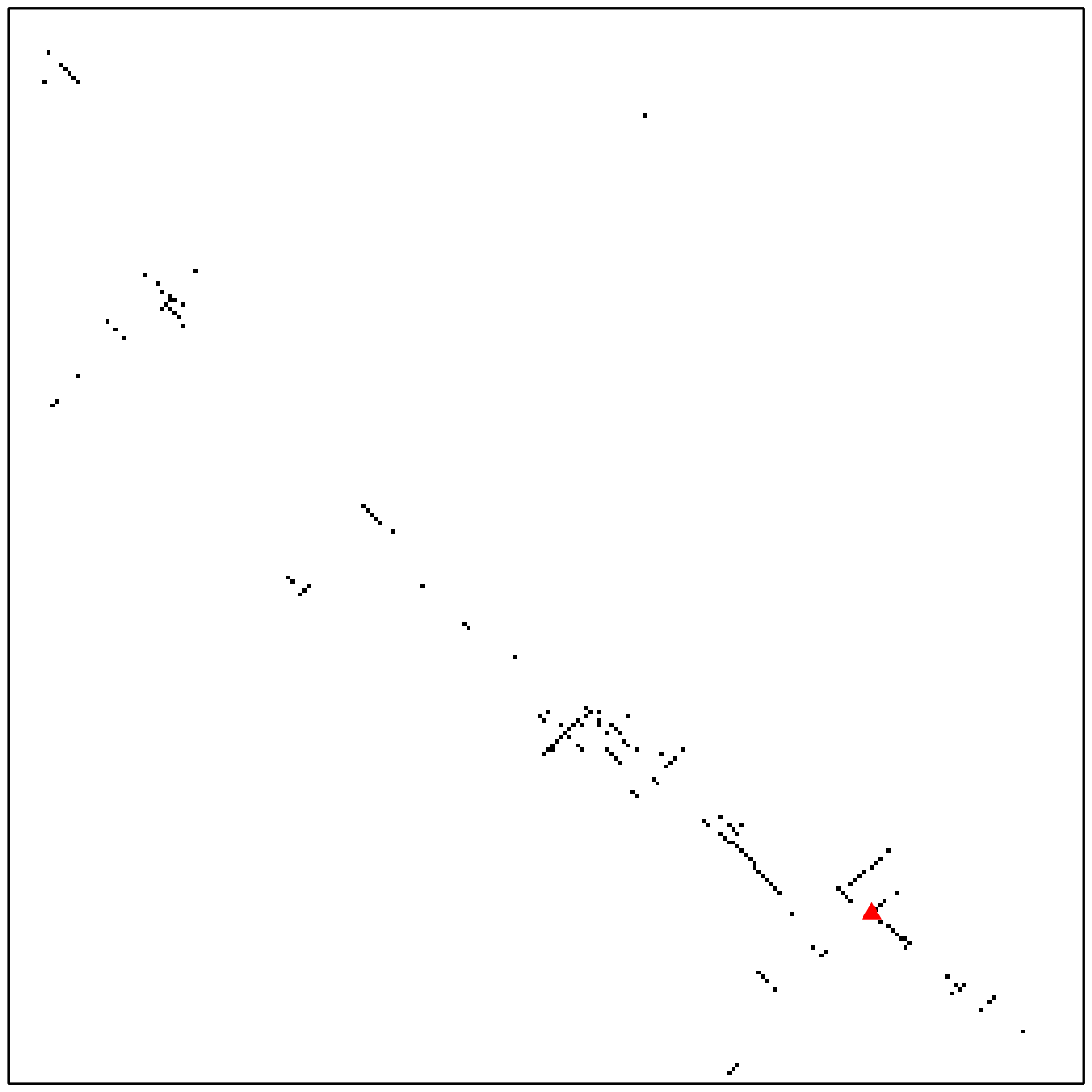}
   \includegraphics[width=57mm]{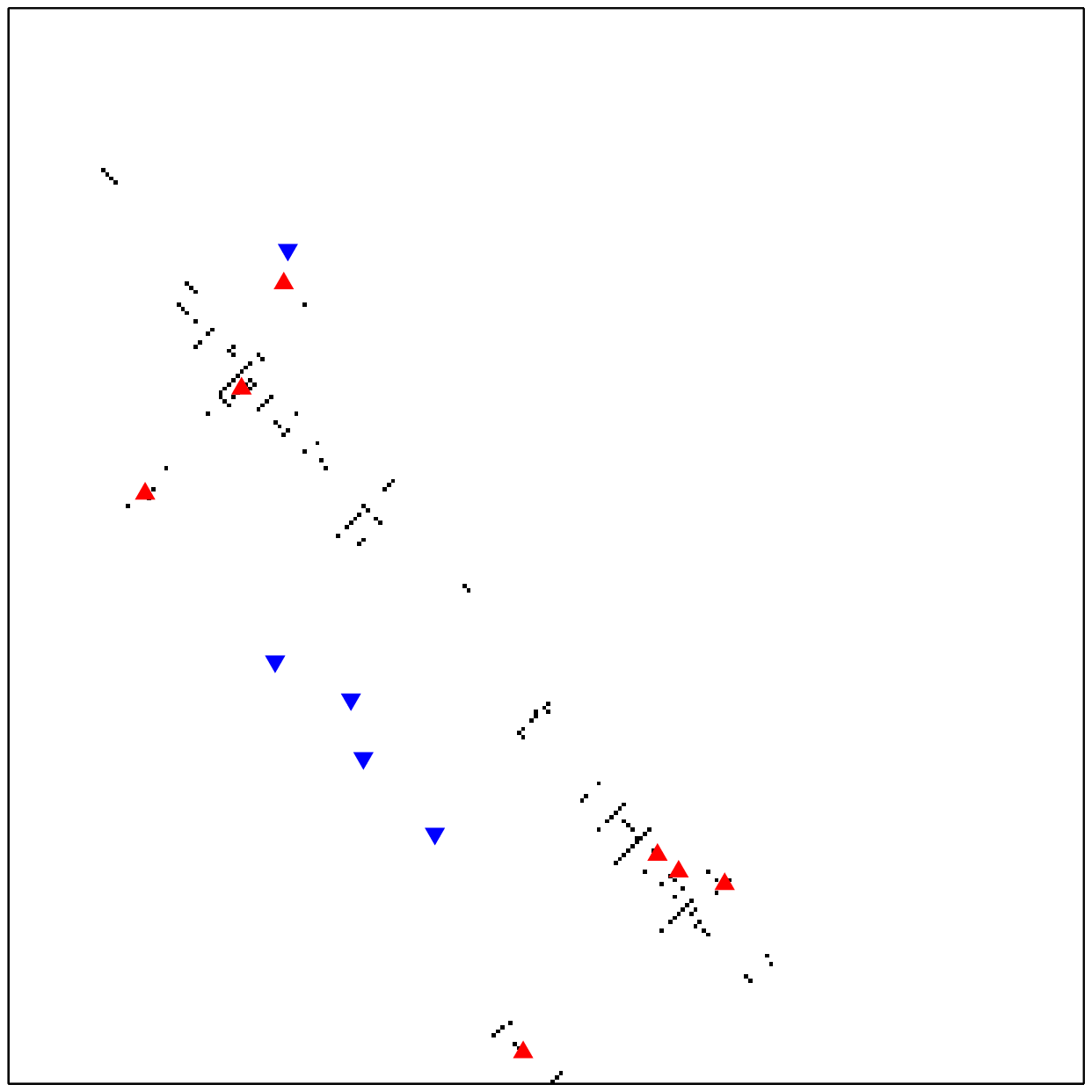}
   \includegraphics[width=57mm]{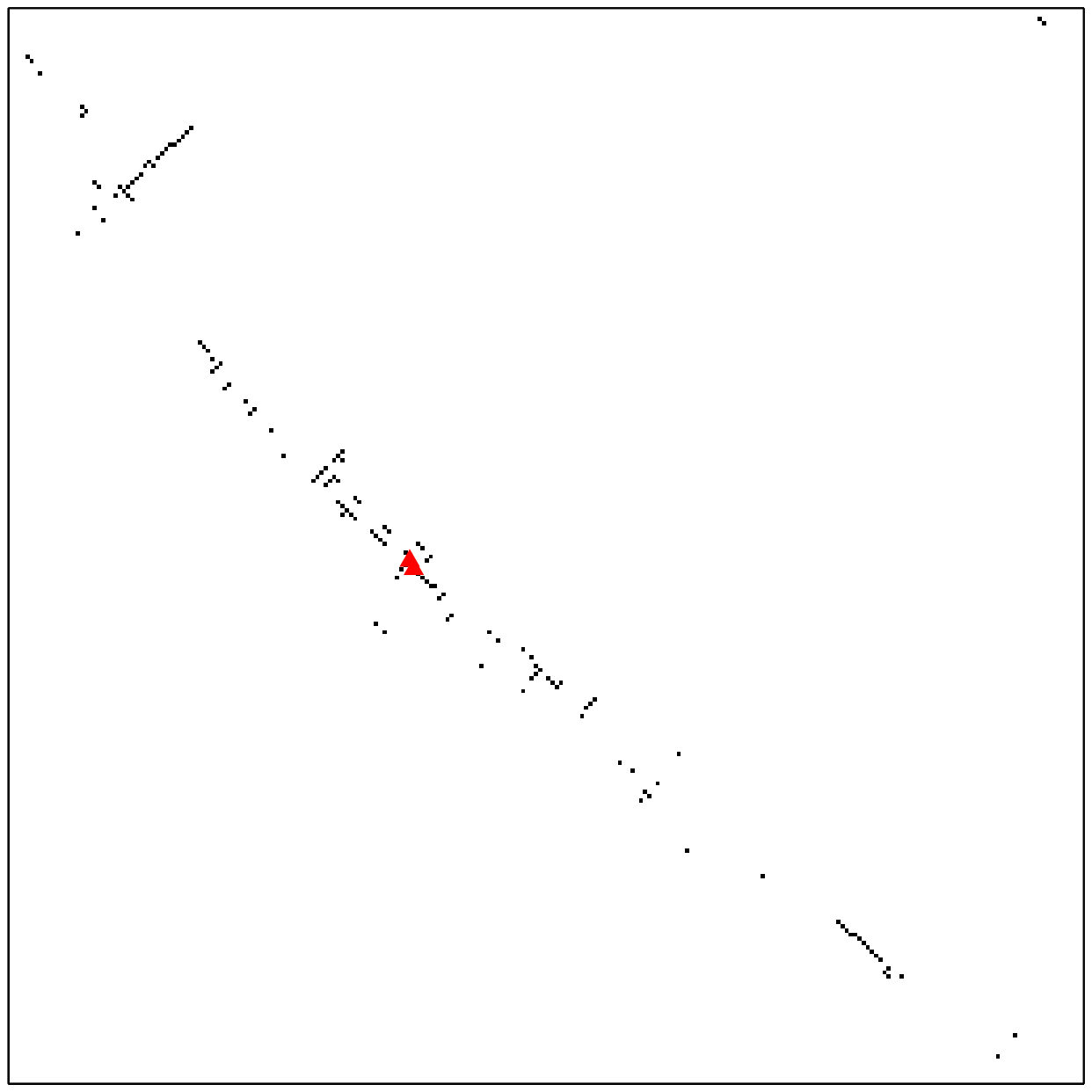}
   \includegraphics[width=57mm]{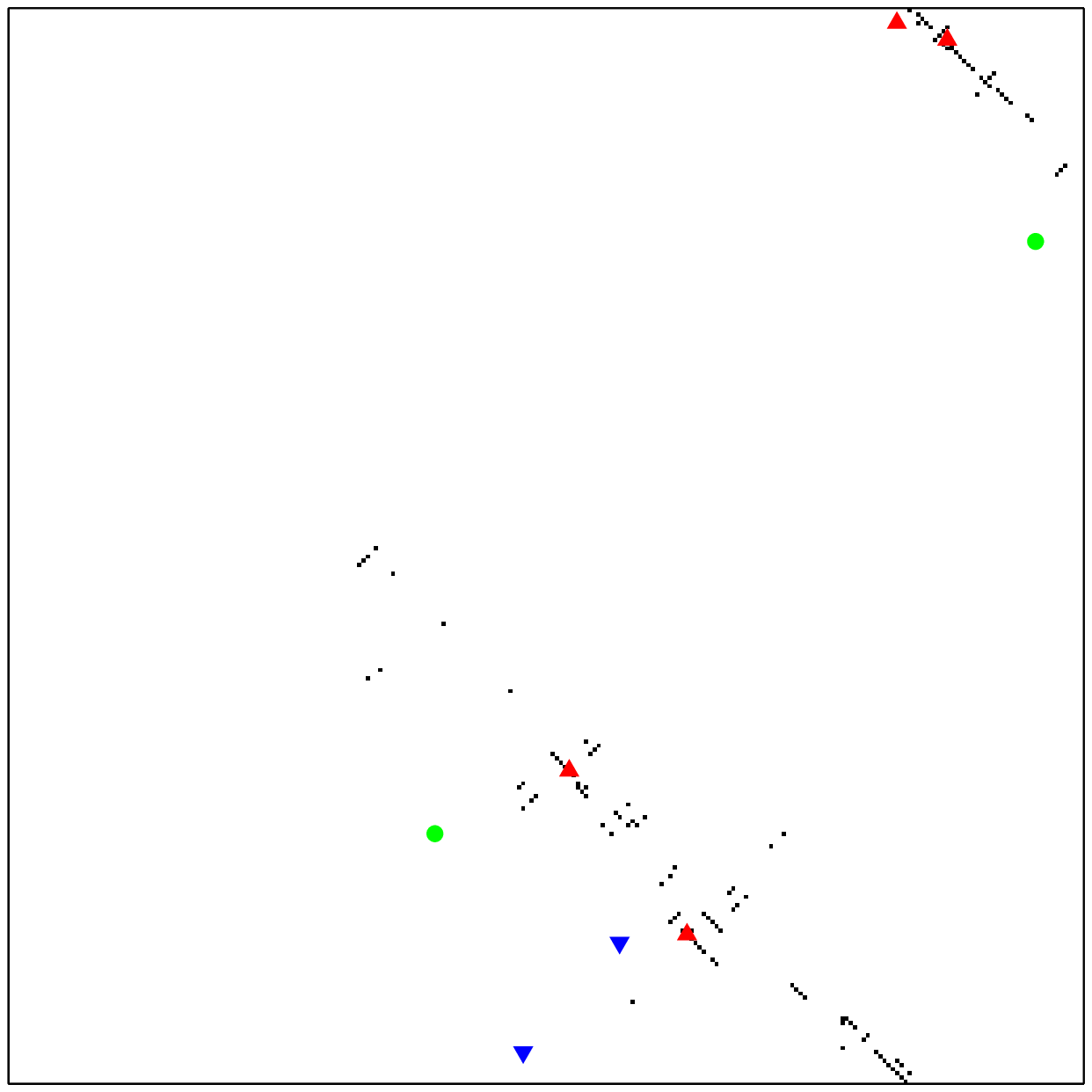}
   \includegraphics[width=57mm]{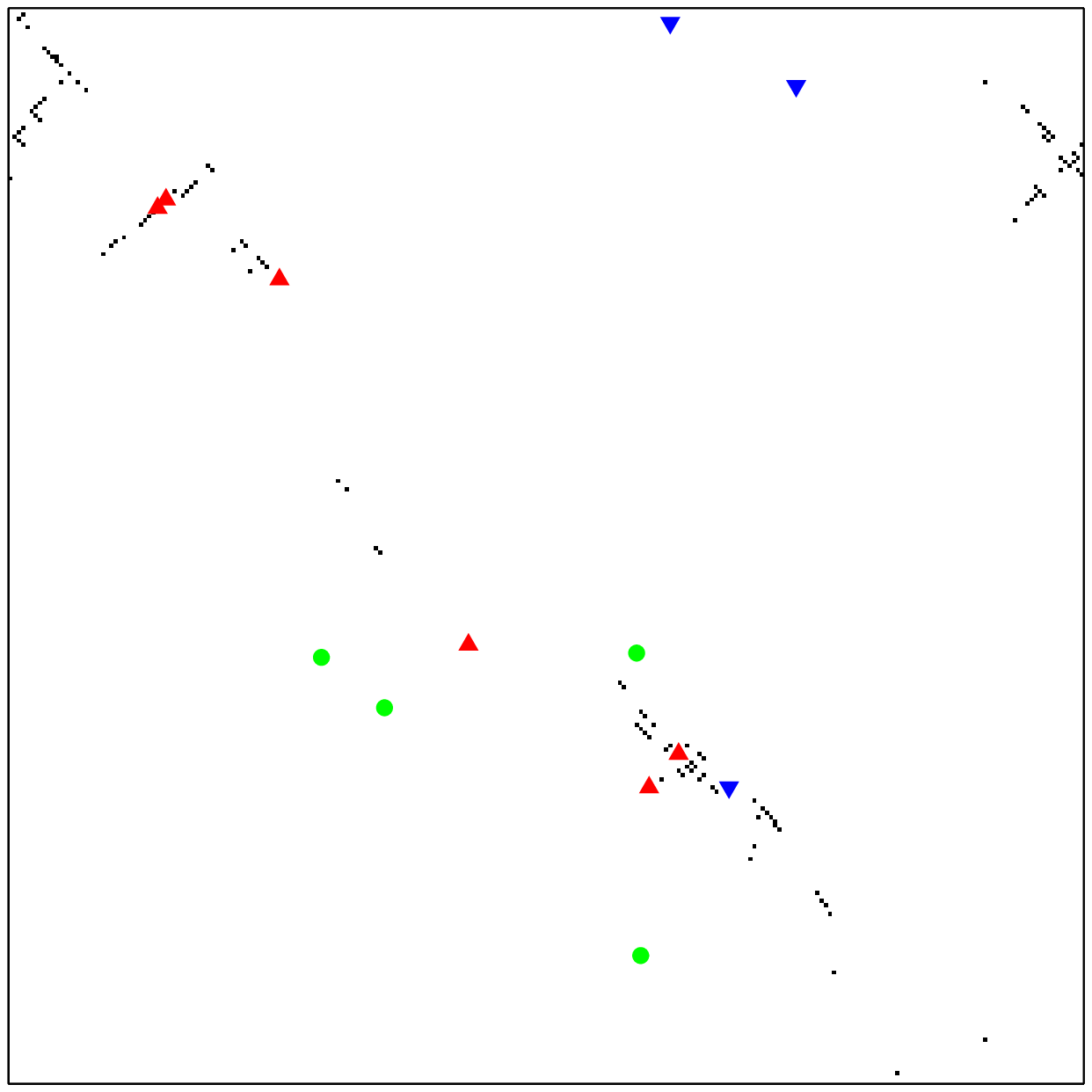}
  \caption{(Color online) Maps of cumulated plastic activity during avalanches
    obtained with stiffness values $K=1$ for a system of size
    $L=256$. The colored symbols indicate the location of precursors just
    before the avalanche takes place. Red upward triangles are part of the
    avalanche, green and blue symbols are not part of the avalanche but
    at the end of the latter green dots are still overcritical while
    blue downward triangles are no longer over-critical, they have been healed
    during the avalanche.}
  \label{Map-Precursor}
\end{figure}

The identification of a set of precursors can be illustrated
graphically. In Fig. \ref{Map-Precursor} we superimposed the plastic
activity observed during avalanches with the set of identified
precusors (represented as colored symbols) just before the avalanche take
place. One can clearly see that avalanches indeed initiate from some
of these over-critical sites. One also observes a striking
intermittence of this population which can fluctuate from one to a few
tens.  It is of interest to follow the fate of these precursors during
the avalanche. The red upward triangle indicate sites taking part in the
avalanche. Blue and green symbols indicate sites not taking part of the
avalanche but at the end of the latter, green dots are still
over-critical while blue downward triangles are no longer over-critical. This
possibility of healing is a specificity of the present model. Indeed,
in contrast to the case of a depinning front where elastic coupling
does not change sign, the quadrupolar interaction is positive or
negative depending on the direction and thus has either a stabilizing
effect (over-critical sites are sent back in the sub-critical part of
the distribution) or a destabilizing effect. However, the presence of
greens dots indicate that not all overcritical sites are exhausted
during an avalanche. Only late plastic events will be initiated there.
A large part of the dynamics of the model is thus related to this
population of precursors which seems to encode a long term
information.

\section{Conclusion}\label{sec:conclusion}

The present meso-model of amorphous plasticity, based on the competition between
a local yield stress randomness and long range elastic interaction allowed us to
obtain non trivial results about avalanche statistics and finite-size effects.
In particular the exponent reported here for the scale free avalanche
distribution, $\kappa\approx 1.25$ is significantly different from the mean
field value (3/2). This suggests that a faithful account of the elastic stress
redistribution due to local restructuring is indeed a crucial ingredient in the
modeling of amorphous plasticity to capture faithfully collective effects.
Although original due to this quadrupolar interaction, the present model can
still be discussed in the framework of the depinning transition. This allowed us
to track the finite size fluctuation and systematic size effect of the
macroscopic yield stress. In addition, a set of precursory sites of having a
fractal support has been identified.

%\bibliography{vdb,plasticity,silica,depinning}

\end{document}